\documentclass[final,3p,times]{elsarticle}

\usepackage{amssymb}
\usepackage{amsmath}

\journal{MethodsX}

\usepackage{longtable}
\usepackage{multirow}
\usepackage{xcolor}

\usepackage{hyperref}

\begin{document}

\begin{frontmatter}



\title{Calculating incidence of Influenza-like and COVID-like symptoms from Flutracking participatory survey data}

\affiliation[inst1]{organization={COVID Modelling Aotearoa, The University of Auckland},
            addressline={38 Princes Street, Auckland CBD}, 
            city={Auckland},
            postcode={1010}, 
            country={New Zealand}}

\affiliation[inst2]{organization={Te P\={u}naha Matatini, The University of Auckland},
            addressline={38 Princes Street, Auckland CBD}, 
            city={Auckland},
            postcode={1010}, 
            country={New Zealand}}

\affiliation[inst3]{organization={M.E. Research},
            addressline={Takapuna}, 
            city={Auckland},
            postcode={0622}, 
            country={New Zealand}}

\affiliation[inst4]{organization={Department of Physics, The University of Auckland},
            addressline={38 Princes Street, Auckland CBD}, 
            city={Auckland},
            postcode={1010}, 
            country={New Zealand}}

\affiliation[inst5]{organization={Department of Engineering Science, The University of Auckland},
            addressline={70 Symonds Street, Grafton}, 
            city={Auckland},
            postcode={1010}, 
            country={New Zealand}}

\author[inst1,inst2,inst3,inst4]{Emily P. Harvey\corref{cor1}}
\ead{ehar061@uoa.auckland.ac.nz}
\cortext[cor1]{Corresponding Author}

\author[inst1,inst4,inst5]{Joel A. Trent}
\author[inst1,inst4]{Frank Mackenzie}
\author[inst1,inst2,inst4]{Steven M. Turnbull}
\author[inst1,inst2,inst4]{Dion R.J. O'Neale}

\begin{abstract}

This article describes a new method for estimating weekly incidence (new onset) of symptoms consistent with Influenza and COVID-19, using data from the Flutracking survey.
The method mitigates some of the known self-selection and symptom-reporting biases present in existing approaches to this type of participatory longitudinal survey data. 

The key novel steps in the analysis are: 

\textbf{1)} Identifying new onset of symptoms for three different Symptom Groupings: COVID-like illness (CLI1+, CLI2+), and Influenza-like illness (ILI), for responses reported in the Flutracking survey.

\textbf{2)} Adjusting for \emph{symptom reporting bias} by restricting the analysis to a sub-set of responses from those participants who have consistently responded for a number of weeks prior to the analysis week.  

\textbf{3)} Weighting responses by age to adjust for \emph{self-selection bias} in order to account for the under- and over-representation of different age groups amongst the survey participants. This uses the \texttt{survey} package \cite{survey_package_r} in R \cite{r_core_team}. 

\textbf{4)} Constructing 95\% point-wise confidence bands for incidence estimates using weighted logistic regression from the \texttt{survey} package \cite{survey_package_r} in R \cite{r_core_team}.

In addition to describing these steps, the article demonstrates an application of this method to Flutracking data for the 12 months from 27th April 2020 until 25th April 2021. 
\end{abstract}




\end{frontmatter}



\section{Method details}
\label{sec:methods}



This article describes a method that can be used to estimate the onset of new incidence of Influenza-like (ILI) and COVID-like (CLI) symptoms from longitudinal participatory health survey data.
The method mitigates some of the known biases in existing approaches for this type of data and provides more robust estimates of symptom incidence over time.
In this article we also demonstrate an application of the methodology to Flutracking data for the 12 months from 27th April 2020 until 25th April 2021 in Aotearoa New Zealand.

As part of this work, we define a range of \emph{Symptom Groupings} that can be used to estimate the incidence of symptoms consistent with COVID-19, and demonstrate the impact of the chosen Symptom Groupings on resulting estimates. Because multiple illnesses can lead to overlapping sets of symptoms, it is important to note that this method does not attempt to diagnose new onset of a specific disease (e.g. COVID-19); rather it estimates the onset of new (sets of) symptoms that are consistent with certain diseases, including COVID-19.

The method and associated code presented here have been developed to analyse data from the Flutracking survey \cite{flutracking_info} in Aotearoa New Zealand, but can be easily adapted to similar longitudinal data, where self-selection biases and reporting biases are known issues.


\subsection{Background} 

Flutracking \cite{flutracking_info} is a participatory health surveillance system for Australia and Aotearoa New Zealand that seeks to estimate the proportion of the population with new onset of symptoms consistent with influenza \cite{dalton2009flutracking}, and more recently symptoms consistent with COVID-19. The survey was developed by Hunter Population Health in collaboration with The University of Newcastle. The survey is administered online to volunteer participants from the public. Registered participants receive a weekly email asking them to report any cold, flu, or COVID-like symptoms experienced in the previous week. The survey also allows for participants to report on behalf of family members in their household, such as young children. 

Participatory health surveillance systems like Flutracking are a valuable resource for tracking outbreaks of infectious disease. They allow for analysis and reporting on how new onset of symptoms differ over time and between seasons, as well as across different geographic regions. However, there are several typical limitations that can reduce the utility of the results that they provide. 

Previous studies have shown that participatory health surveillance systems can be biased in terms of survey participants and response rates over time. Firstly, some participants may be more likely to respond to a survey if they have experienced symptoms in a given week\cite{Baltrusaitis2018,Liu2020} (a form of \emph{reporting bias}), and secondly, participatory samples rely on volunteers and are not representative of the general population \cite{Friesema2009_Internet_based_monitoring, Debin2013} (a form of \emph{self-selection bias}). In particular, people from lower socio-economic groups and minority ethnicities are found to be under-represented \cite{Scarpino2020_Socioeconomic_bias_influenza}. Within Aotearoa New Zealand, we find that Flutracking respondents are disproportionately of P\={a}keha/NZ European ethnicity and older, with M\={a}ori and Pacific Islanders, and people under 30 years old, being particularly under-represented (see Section \ref{sec:demographic}). Both of these biases should be accounted for where possible to improve the reliability of incidence estimates \cite{Cantarelli2014,Brownstein2017_Combining_Participatory_Influenza}.

Participatory health surveillance studies have commonly focused on influenza tracking and have relied on strict symptom definitions that only describe one experience of influenza-like illness (ILI). For example, previous studies, including current weekly Flutracking reports \cite{flutracking_june2022}, focus solely on experiences of ILI defined as new onset of both cough and fever in the same week \cite{Chunara2015}. While these symptoms are commonly viewed as core components of ILI \cite{dalton2009flutracking}, this criteria may be insufficient for investigating the rates of a wider range of respiratory illnesses, and in particular, would miss the majority of COVID-19 cases \cite{Lavezzo2020Vo,Wells2020symptoms,SymptomClusters}.

In the context of Aotearoa New Zealand, during the year of data considered in this article (27th April 2020 to 25th April 2021), the COVID-19 Elimination Strategy advised anyone who experienced symptoms consistent with COVID-19 to seek a test, so as to promptly detect and stamp-out any new community outbreak of SARS-CoV-2. Monitoring the performance of such an approach requires estimating the incidence of symptoms that are consistent with COVID-19, including those that do not meet the strict `cough and fever' ILI definition, in order to determine whether testing rates in a given week are commensurate with incidence of new COVID-like symptoms in the same week.

In this article, we describe a methodology that seeks to address some of the limitations described above for data from the Flutracking survey from Aotearoa New Zealand. Despite this specific application, the method can be applied equally to Flutracking data from Australia, or to similar longitudinal data collected elsewhere (e.g., the ``Flu Near You'' survey in the United States \cite{smolinski2015flu}, or  ``Influenzanet'' in Europe \cite{Paolotti2014}). Our approach for determining a subset of responses that are classified as `consistent' in any given week, in order to reduce reporting bias and to maximise response numbers retained, is also applicable to other datasets with similar issues. For example, it could be easily applied to prevalence estimates calculated from `test-to-stay' type surveillance data COVID-19 from Rapid Antigen Test results by filtering to only consider results from individuals who were regularly reporting (negatives) in the weeks leading up to a positive test.

\subsection{Flutracking data}
Raw data are obtained from Flutracking New Zealand \cite{flutracking_info} through the New Zealand Ministry of Health. 
Flutracking symptom data consists of a presence/absence indication for each respondent for each of six\footnote{A seventh symptom --- headache --- was added to the survey in the second half of 2021, to capture better the reported symptoms for the Delta variant of SARS-CoV-2. We do not include Headache in our analysis, even after it was added, in order to maintain consistency.} symptoms: cough, fever, sore throat, shortness of breath, runny nose, and loss of taste or smell. All of these symptoms are part of the New Zealand Ministry of Health's definition of `COVID-like symptoms' \cite{MoHCovidSymptoms}. 

\begin{figure}[htb!]
    \centering
    \includegraphics[width=0.7\textwidth]{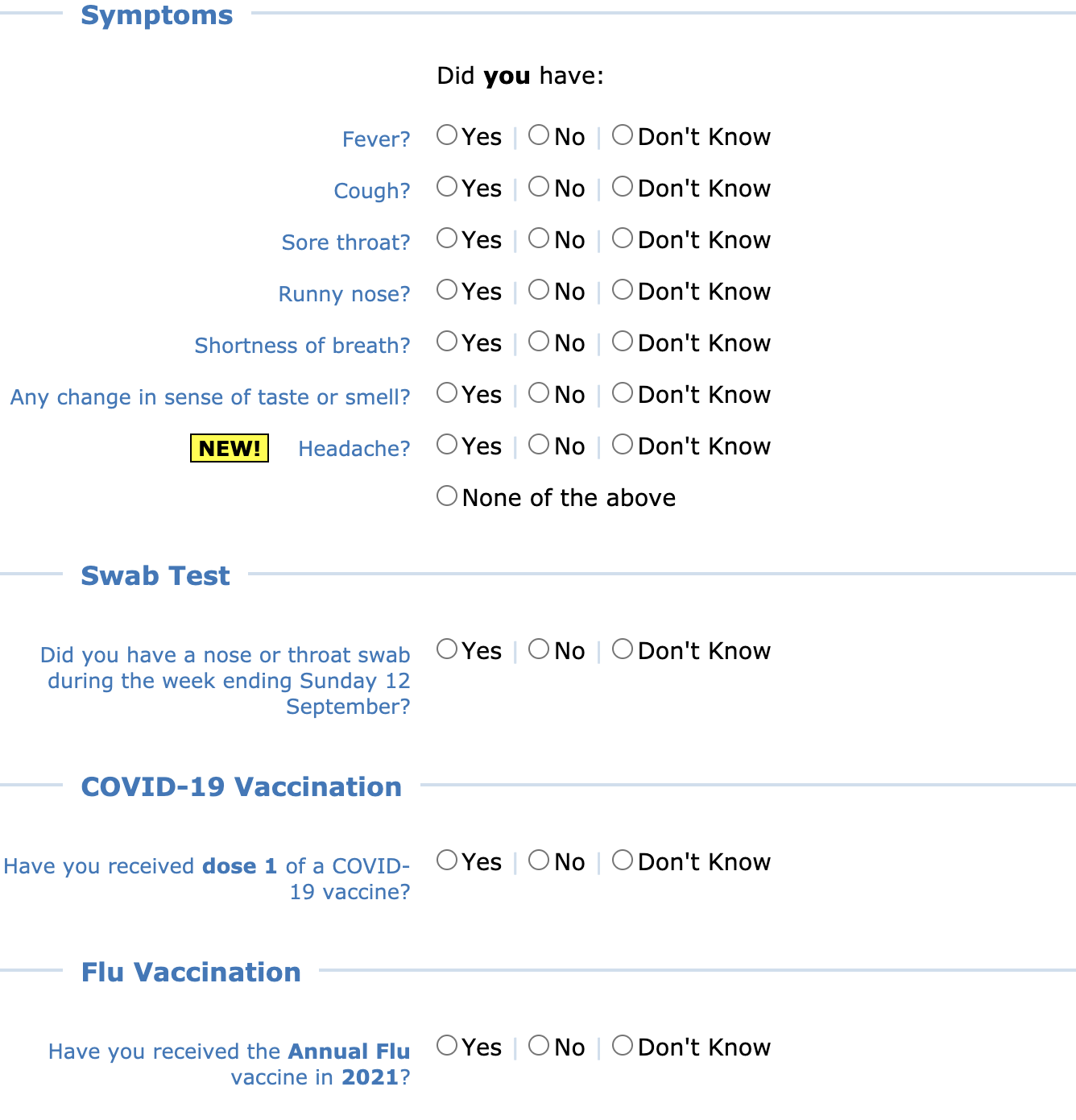}
    \caption{Example of the Flutracking weekly survey, sent to participants via email. Retrieved from \url{https://www.flutracking.net/Demo/NZ} \cite{flutracking_info}}
    \label{fig:survey}
\end{figure}

When respondents initially enrol in the Flutracking survey, they are asked to give demographic information, including age, ethnicity, location (postcode), and gender. Participants are also asked about their vaccination status for COVID-19 and the annual influenza vaccine, and whether they have been tested for COVID-19 in the previous week. An example of the weekly survey is shown in Figure~\ref{fig:survey}. If a respondent indicates that they are experiencing symptoms, they are then asked whether they sought healthcare or missed work/usual activities due to the illness, and whether they were tested for COVID-19 or Influenza.

\subsection{Method outline}
For each survey week in a period of interest, the following steps are applied:
\begin{itemize}
    \item For each response that matches the \emph{Symptom Grouping} of interest, classify whether that week is a \emph{new onset} of those symptoms.
    \item Determine which responses each week count as \emph{consistent responses}, and remove those that do not.
    \item Calculate an \emph{age weighting factor} for each age group within the survey population, to match the age distributions of the consistent responses each week with a known population distribution. This is done using the \texttt{rake} function within R's \texttt{survey} package \cite{survey_package_r}. The age weighting factor is calculated for each week, and can be calculated at either a whole-country level or for sub-populations specified as sets of locations and ages.
    \item Construct \emph{95\% pointwise confidence bands} around mean incidence estimates using the \texttt{svyciprop} function in R's \texttt{survey} package for specified sets of factors, including age group, Symptom Grouping, location, and survey week.
\end{itemize}
The following sections detail each of the steps of the method in more depth.

\subsubsection{Define Symptom Grouping criteria}
We define three different Symptom Groupings: 
\begin{itemize}
    \item \textbf{CLI1+}: Responses indicate any one or more of the above \textit{COVID-like} symptoms. CLI1+ meets the Aotearoa New Zealand Ministry of Health advice to seek a COVID-19 test\cite{MoHCovidSymptoms}.
    \item \textbf{CLI2+}: Responses indicate two or more of the above COVID-like symptoms. CLI2+ allows us to be slightly more discerning, given that many non-infectious illnesses such as allergies or asthma can cause a new onset of the COVID-like symptoms in the survey. 
    \item \textbf{ILI}: Responses indicate at least both \textit{cough} and \textit{fever}, which are symptoms of \textit{influenza-like illness}. This Symptom Grouping is used for public Flutracking reporting \cite{flutracking_info}.  
\end{itemize}

\subsubsection{Incidence classification}
We wish to estimate `incidence', not `prevalence' from the Flutracking data. Specifically, we wish to identify any \textit{new} onset of symptoms that meet the threshold for an `incident' according to a chosen Symptom Grouping. The method we use follows the same method employed by the Aotearoa New Zealand Ministry of Health and Flutracking Australia for estimating the incidence of ILI (cough and fever), but extends to the Symptom Groupings CLI1+ and CLI2+ above, in addition to the Symptom Grouping for ILI.

\begin{figure}[htb!]
    \centering
    \includegraphics[width=\textwidth]{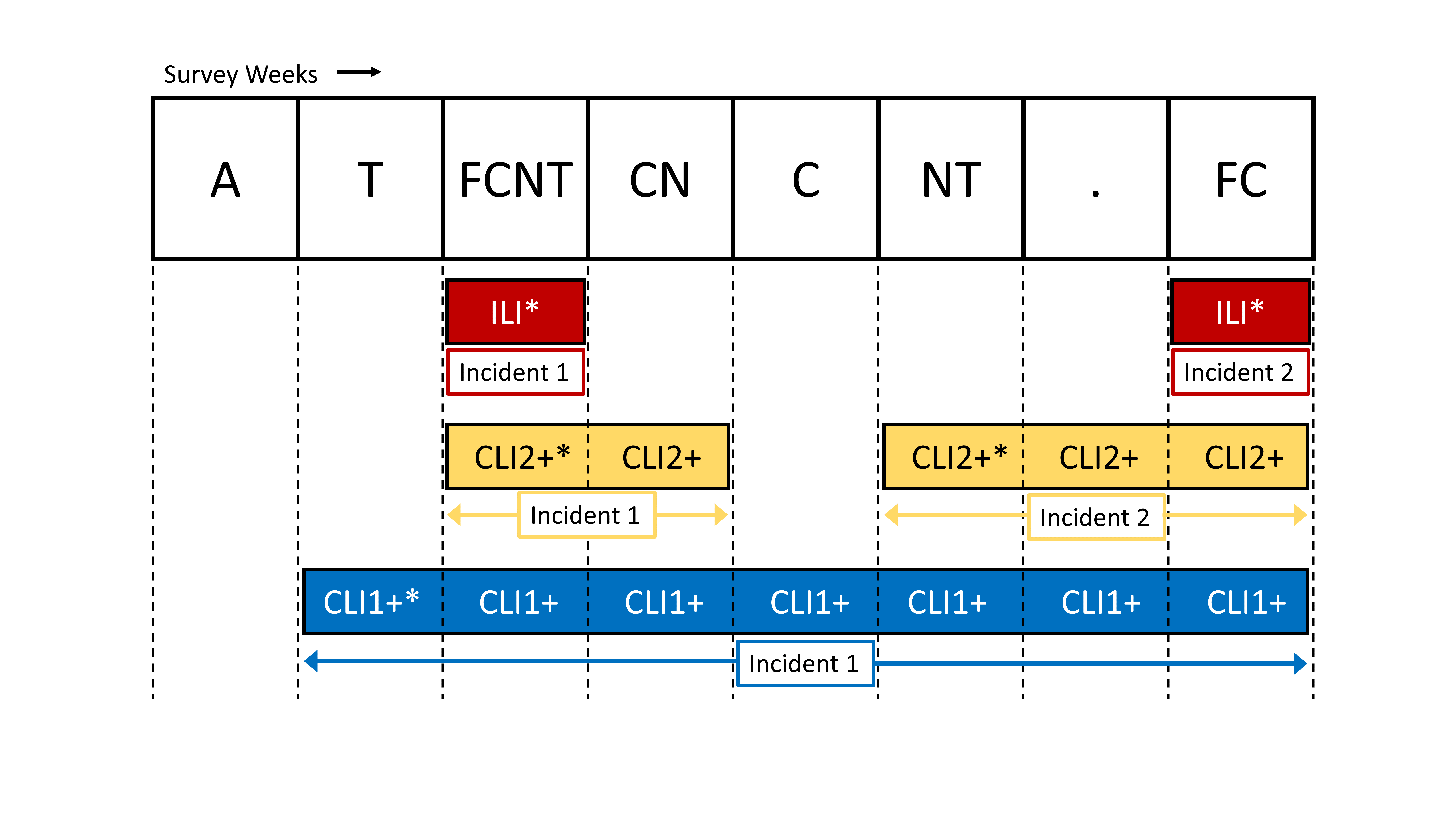}
    \caption{Example of symptoms reported each week for one hypothetical respondent. This illustration shows a period of 8 survey weeks, with reported symptoms (or non-response) indicated for each week. The symptoms reported include \textbf{A} Asymptomatic (i.e. a response indicating no symptoms), \textbf{C} Cough, \textbf{F} Fever, \textbf{N} runny Nose, and \textbf{T} sore Throat. A dot indicates the respondent did not answer the survey that week. In this hypothetical case, we see how the responses given can fall under each Symptom Grouping and that these \textit{Incidents} cover different survey weeks depending on the patterns of symptoms. As CLI1+ is the most inclusive Symptom Grouping, the criteria for this are met in most weeks shown here and it is considered a single \textit{Incident}, while the Symptom Groupings ILI and CLI2+ would be split into two separate \textit{Incidents} during the same time period. We can also see that CLI1+ and CL2+ (but not ILI) span a period including one week where no response was given. This is because we assume that the missing week would be a continuation of the same \textit{Incident}, provided it meets the same Symptom Grouping criteria. The * indicates which week will be included as the onset of the \textit{Incident}, and will be included in the `Incidence' calculations.}
    \label{fig:Incidence_example}
\end{figure}

Flutracking data contains each respondent's reported symptoms for a given survey week. For a selected Symptom Grouping, we first check whether an individual's symptoms in the given week meet the criteria for that Symptom Grouping. If so, we mark it as an Incident. For example, if we wanted to investigate the number of CLI1+ Incidents, we would record all weeks where an individual recorded one or more of the six COVID-like symptoms as an Incident. 

Figure~\ref{fig:Incidence_example} provides a hypothetical example of a respondent who reported different symptoms over the course of 8 weeks.

To distinguish new onset of illness from the continuation of a previously reported Incident, each Incident is assigned a unique ID, with this same Incident ID given to all consecutive weeks that meet the Symptom Grouping. If there is no response for one week between two weeks in which the Symptom Grouping was met, it is assumed that the interstitial week is a continuation of the previous week's Incident and it is allocated the same Incident ID. For any gaps in responses between Incidents longer than one week we allocate the second Incident a new Incident ID. This procedure is consistent with the existing methodology used by Flutracking Australia and Flutracking New Zealand.

The incidence calculation is applied independently for each Symptom Grouping. This means that if a respondent reported one symptom in their first week of illness and two or more symptoms in the second week, the second week would be recorded as an `Incident' of CLI2+, as though it were the onset of a new illness. This means that if a participant met the criteria for CLI1+ for three consecutive weeks, but met the criteria for CLI2+ only in the first and last weeks, this would be recorded as two Incidents of CLI2+, as shown in Figure \ref{fig:Incidence_example}. This is capturing useful information, as the advice is that someone with new or worsening symptoms should seek a COVID-19 test.

While our approach allows for flexible Symptom Grouping criteria to be implemented within the methodology, future work may seek to apply more sophisticated, unsupervised learning approaches when defining Incidents. While not in the scope of the current research, previous work has demonstrated that unsupervised learning approaches provide a useful technique for modelling heterogeneous symptom experiences among respondents \cite{Kalimeri2019}.


\subsubsection{Determining the set of consistent responses}

Flutracking participants are more likely to respond to the survey when they are experiencing symptoms than when they are well\cite{Liu2020}. This \emph{symptom reporting bias} can lead to overestimating symptom incidence rates if it is not accounted for. Consequently, it is important to define a subset of \emph{consistent responses} to use for incidence rate calculations. These are defined as being the responses, in a given week, from participants who are deemed to have responded consistently to the survey, as assessed at that week.

Previous studies have determined consistent respondents by only considering responses from those participants who have completed more than some minimum fraction of all surveys (e.g.,~\cite{Chunara2015,Debin2013}). We improve on this approach by determining a subset of consistent responses for each week of the analysis, based on the response history of each survey participant for a defined window prior to the week under consideration. This allows a respondent's `consistent' status to change from week to week. We implement this by looking back over a specified number of weeks (the `window size') and requiring the participant to have responded in all those weeks, with an allowance for a certain number of `missing weeks' within the window. This makes it possible to adjust for the known reporting bias at any given point in time, while still maximising the sample size at the corresponding time. Our approach also means that analysis can be performed on a weekly basis throughout the year, rather than only at the end of a year (or season) of data collection. 

\begin{figure}[htb]
    \centering
    \includegraphics[width=\textwidth]{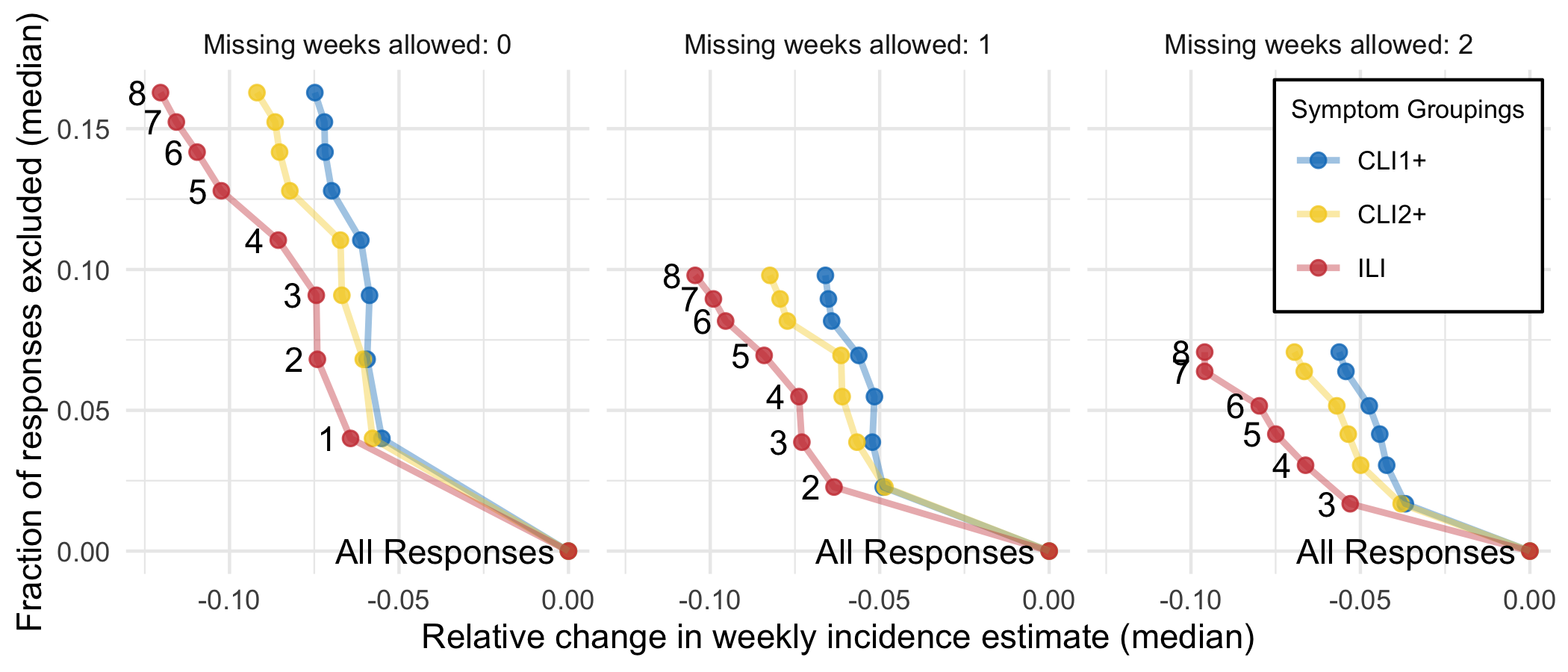}
    \caption{Curves showing the fraction of responses excluded and the relative change in weekly incidence rate estimates for different choices of window size (text labels on the plot) and missing weeks allowed (panels left to right), for each of the three Symptom Groupings (colour of points and lines).}
    \label{fig:Consistent_tradeoffs}
\end{figure}

In Figure~\ref{fig:Consistent_tradeoffs} we plot trade-off curves showing the median fraction of responses excluded and the median relative change in the weekly incidence estimates as a function of `Window size' and `Missing weeks allowed'. This helps us to determine suitable parameter values for the selection criteria of consistent responses. For more information on how these effects vary throughout the year, see Figures~\ref{fig:responsesChange} and \ref{fig:incidenceChange} in the Additional Information section.

Increasing the number of prior weeks for which a participant must have previously responded (the response window size) reduces the number of participants  included in the `consistent response' subset for any given week --- that is, the fraction of responses excluded increases with increasing window size. Allowing respondents to have skipped some of the weeks in the consistent response window means that the responses of some participants are re-included in the consistent response subset.

When we consider the impact of window size and missing weeks allowed on the incidence estimate we find that a more stringent (i.e.~larger) window size reduces the rate of symptom incidence calculated for any given week. This is expected from previous work looking at the impact of reporting bias \cite{Liu2020}. The biggest jump is seen when requiring \emph{any} response in the window prior to reporting symptom onset (i.e.~a one-week window, or a two-week window with one missing week allowed, or a three-week window with two missing weeks allowed). Imposing this minimal consistent response window results in an immediate decrease of approximately 5\% in the incidence estimate for all Symptom Groupings. We attribute this initial decrease to a correction for the known symptom reporting bias --- if participants were equally likely to respond in a given week, independent of symptom incidence, then applying this criterion would not affect the estimated incidence rate.

Increasing the consistent response window, beyond the minimum, initially causes a decrease in the number of consistent responses but does not significantly change incidence estimates. However, further increasing the window size beyond about four weeks causes both the number of consistent responses and the estimated incidence rate to decrease. We attribute this to the more stringent consistent response window introducing a different form of bias by increasing the over-representation of some groups of participants in the sample population --- specifically older P\={a}keh\={a}/NZ European cohorts who tend to have lower incidence rates --- see for example Figure~\ref{fig:age_incidence}.

\begin{figure}[ht]
    \centering
    \includegraphics[width=\textwidth]{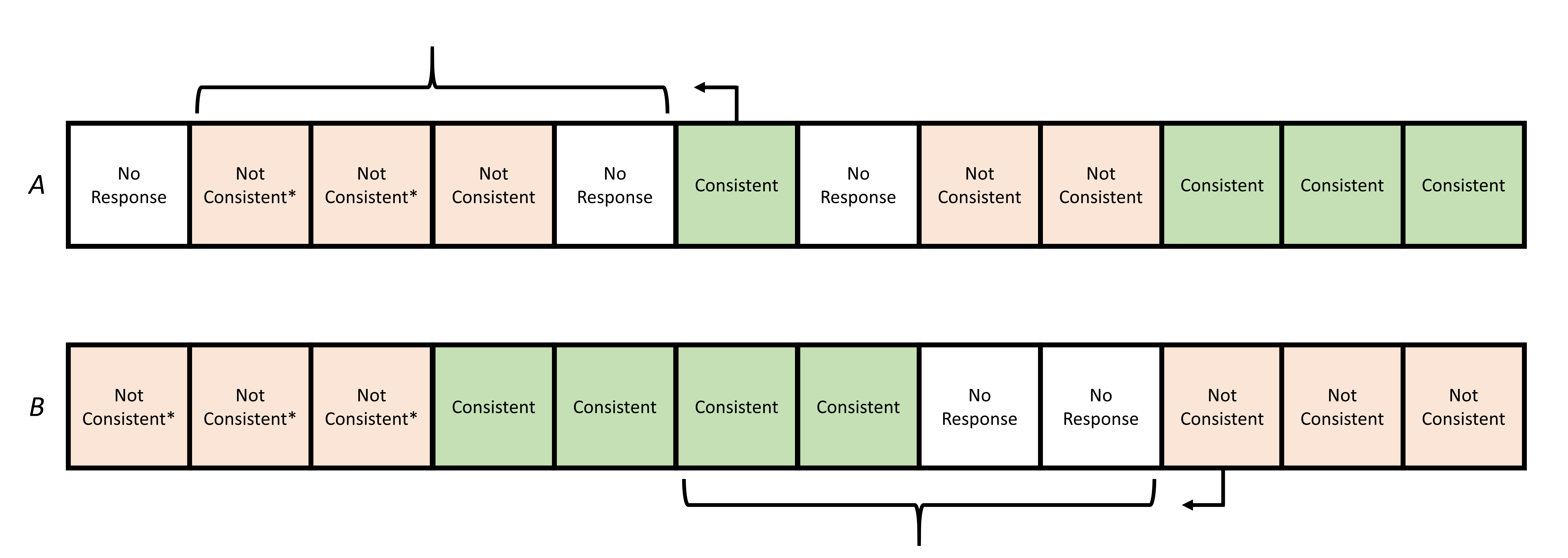}
    \caption{Hypothetical examples for the selected window size of `the previous 4 weeks', and allowing 1 missing response in that window. The illustration above shows survey response series for hypothetical respondents \textit{A} and \textit{B}. Any weeks with `no response' are left uncoloured. For weeks with a response, that response is defined as `not consistent' (pink), or `consistent' (green). We can see that for respondent \textit{A}, the highlighted survey week is considered `consistent' as 3 of the previous 4 survey weeks contain a response (even if for each of those previous survey weeks, a response was considered not consistent). For respondent \textbf{B}, we can see that the highlighted survey week is considered `not consistent' because 2 out of the previous 4 survey weeks did not contain a response. * indicate that the survey weeks covering the respondents first three weeks of participation are considered not consistent, due to the fact that they have not yet answered enough surveys to meet the threshold of consistency.}
    \label{fig:Consistent_responses_example}
\end{figure}

While our method allows for the user to choose any consistent respondent window size and any number of missed weeks within that, we have chosen in our subsequent analysis to use a window size of four weeks with up to one missing week. This defines a consistent response subset where small changes in the window size do not result in significant changes in the resulting incidence rate and where the fraction of responses included in the sample population for a given week is not too severely reduced, in part due to the fact that the Flutracking survey allows respondents to retrospectively enter results for up to four weeks into the past.
We provide two examples of the selected values for window size and allowed missing responses using hypothetical response series illustrated in Figure \ref{fig:Consistent_responses_example}.

\subsubsection{Accounting for self-selection biases: age weighting}\label{sec:age_weighting}

There are several demographic biases in the Flutracking cohort with \emph{registration rates} differing by age, ethnicity and location of respondents. If any of these self-selection biases also align with a corresponding variation in \emph{incidence rates} along the same demographic factor, then it will contribute to an under- or over-estimate of incidence for the total population. 


\begin{figure}[htb!]
    \centering
    \includegraphics[width=0.6\textwidth]{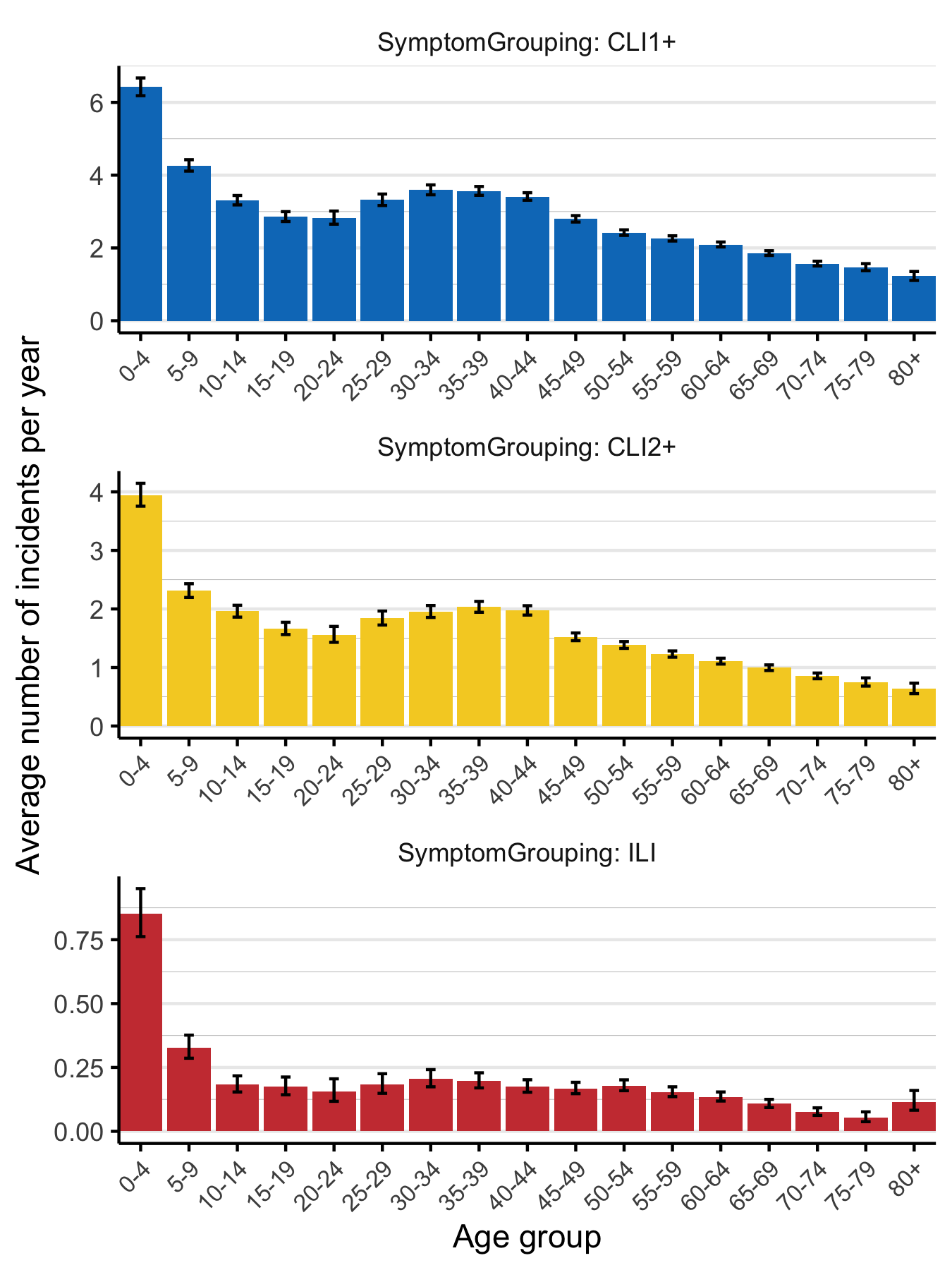}
    \caption{Average number of new incidents per person for the year between 27th April 2020 and 25th April 2021, by five-year age groups (error-bars represent 95\% confidence intervals). This only considers respondents who submitted at least 50 surveys over the year. In terms of Symptom Groupings, overall we see that incidents of CLI1+ are more common than CLI2+ and ILI. We can also see that the highest level of incidence, across all three Symptom Groupings, is experienced by those in the 0--4 year old age group. }
    \label{fig:age_incidence}
\end{figure}

The strongest factor affecting incidence estimates is age (see Figure \ref{fig:age_incidence}), with younger ages, particularly 0--4 years, tending to have higher incidence across CLI1+, CLI2+, and ILI Symptom Groupings. Therefore, we adjust our incidence estimates to account for the relative number of respondents in the survey population according to age each week.

Although there is a relationship between ethnicity and incidence of several respiratory illnesses in Aotearoa New Zealand \cite{BAKER2012, HealthyHousing}, it can be unhelpful to treat ethnicity as a predictive factor for these illnesses. Research has found that factors including socioeconomic deprivation, household crowding, and housing quality, are all associated with increased incidence of respiratory infections \cite{HealthyHousing,Ingham2019, AsthmaFoundation}, and that these are highly correlated with ethnicity \cite{Ingham2019}. These same underlying factors are also likely to contribute to a participant selection bias within the Flutracking survey. Under-representation of lower socio-economic groups has also been noted within similar participatory surveys for Influenza. \cite{Scarpino2020_Socioeconomic_bias_influenza}

The low response rate in Aotearoa New Zealand Flutracking data for ethnic groups other than P\={a}keha/NZ European, combined with the under-representation of lower socio-economic groups, mean that it is unlikely that the sampled population of non-P\={a}keha ethnicities in Flutracking is truly representative of those groups in the wider population.
Therefore, weighting based on ethnicity alone may actually amplify selection biases rather than mitigating them. For this reason, although it is possible using our method, we do not calculate or re-weight symptom incidence by ethnicity.
We note that the current re-weighting methodology is easily able to be applied to ethnicity or socio-economic status, should Flutracking enrolment be sufficiently broad that responses are representative of these groups. In order to avoid this limitation in future studies, we point to previous research on best practice for participatory surveillance data collection \cite{Dalton2017_Insights_Flutracking_Thirteen,Smolinski2017}.

The reference population used for the age weightings presented in this method is the 2021 Estimated Residential Population, calculated by Statistics NZ with 5-year age groups and spatial units of District Health Board (DHB) of residence \cite{StatisticsNZ2021}. However, any other source of reference population data can be used, provided that it can be mapped on to the age (1-year age bands) and location (Aotearoa New Zealand postcode) groupings used in the Flutracking data. In general, it is better to use higher levels of aggregation (e.g. 5-year age groups) and DHB. This is in order to avoid the low response numbers per analysis unit, and consequent lower statistical power and increased uncertainty, that a more granular aggregation will lead to.


In order to weight survey responses according to participant age, we use the \texttt{rake} function provided in R's \texttt{survey} package \cite{survey_package_r} to assign a weighting coefficient to each response. The process of raking matches the marginal distribution of ages from the survey respondents with those of a reference population.\footnote{Since this is specified in 5-year age groups, sets of ages must also consist of combinations of 5-year age groups.} The raking can be applied at any level of aggregation appropriate to the reported results, with the caveat that fine-grained aggregation can lead to small cell sizes which may limit the statistical power of results in some cases. 

When analysing responses for sub-national spatial units (e.g.~DHBs), we calculate the age weighting factor for each age group with respect to the age distributions for each of the corresponding sub-national units. 
However, for analysis at a national level, we calculate only national level age weightings, rather than combining sub-national weighting factors. This avoids the pit-fall of national level weighting factors inheriting lower statistical power and increased uncertainty from small cell sizes for some location-age combinations.

\subsubsection{Constructing mean incidence estimates with 95\% confidence intervals}

Constructing confidence intervals for weighted survey data is non-trivial. We construct estimates of mean incidence with 95\% confidence intervals for each survey week for a given set of demographic factors and Symptom Groupings using the \texttt{survey} package in R \cite{survey_package_r}, incorporating the age weighting previously computed in Section \ref{sec:age_weighting}. The 95\% confidence intervals for each survey week can be turned into 95\% \emph{pointwise} confidence bands across survey weeks. 

We first create a `survey design' object using our set of consistent responses, their incidence classification and weighting, and other relevant factors such as their survey week, age group, and location. This survey design object is then passed to the \texttt{svyby} function, which independently calls the \texttt{svyciprop} function for each subset in a set of factors and outputs the estimated mean incidence and 95\% confidence intervals for each subset. \texttt{svyciprop} has several methods of producing confidence intervals; we use its logistic regression method (\texttt{"logit"}). An example set of factors could be Symptom Grouping and survey week, as in Figure~\ref{fig:incidence}. One independent subset used in these calculations is the Symptom Grouping `CLI1+' and the Survey Week ending on Sunday the 31st of May 2020. 

The process used by \texttt{svyby} and \texttt{svyciprop} is equivalent to fitting a weighted logistic regression model, where every subset in a set of factors is fitted independently such that the fitted value of any subset in the set of factors is equal to the weighted mean incidence of that subset. Thus the same model could have been fitted using the \texttt{svyglm} function from the \texttt{survey} package and then fitting interaction effects between all factors. In order to do this, a quasi-binomial model family needs to be used in \texttt{svyglm} \cite{survey_package_r} because the use of survey weights may make the number of `Incidents' non-integer. However, as we were mainly concerned about the fitted means and confidence intervals for each subset rather than fitted values of regressors and their significance, we did not do this in general. Additionally, our process makes it very simple to extract confidence intervals for our estimates, which, while not too difficult (e.g.~using a non-parametric bootstrap \cite{efron_better_1987}), is not nearly as straightforward to do with the fitted model returned by \texttt{svyglm}. 

The only exception to this, is that in order to consider the significance of differences in incidence for a given Symptom Grouping, as seen in Figure~\ref{fig:incidence_location}, we did need to use \texttt{svyglm} to test whether the difference in the estimated mean incidence between two locations on a given week was statistically significant (i.e.~if the p-value for the Auckland Metro DHBs coefficient was less than 0.05).



\subsection{Method demonstration} \label{ssec:method_demo}
We demonstrate the method outlined above by applying it to Flutracking data from Aotearoa New Zealand for the period from the survey week ending Sunday 3rd May 2020 to the survey week ending Sunday 25th April 2021. During this period there was no widespread community transmission of SARS-CoV-2 (the virus that causes COVID-19 disease) in Aotearoa New Zealand. The Elimination Strategy the country employed during this period meant that any detection of SARS-CoV-2 in the community was immediately followed by a system of transmission reduction measures known as Alert Levels \cite{AlertLevels}, which included school closures and stay-at-home orders. 

All plots are generated using the \texttt{ggplot2} package \cite{ggplot2} in R\cite{r_core_team}, with code available at \cite{gitlab_repo}.

\begin{figure}[htb!]
     \centering
     \includegraphics[width=0.95\textwidth]{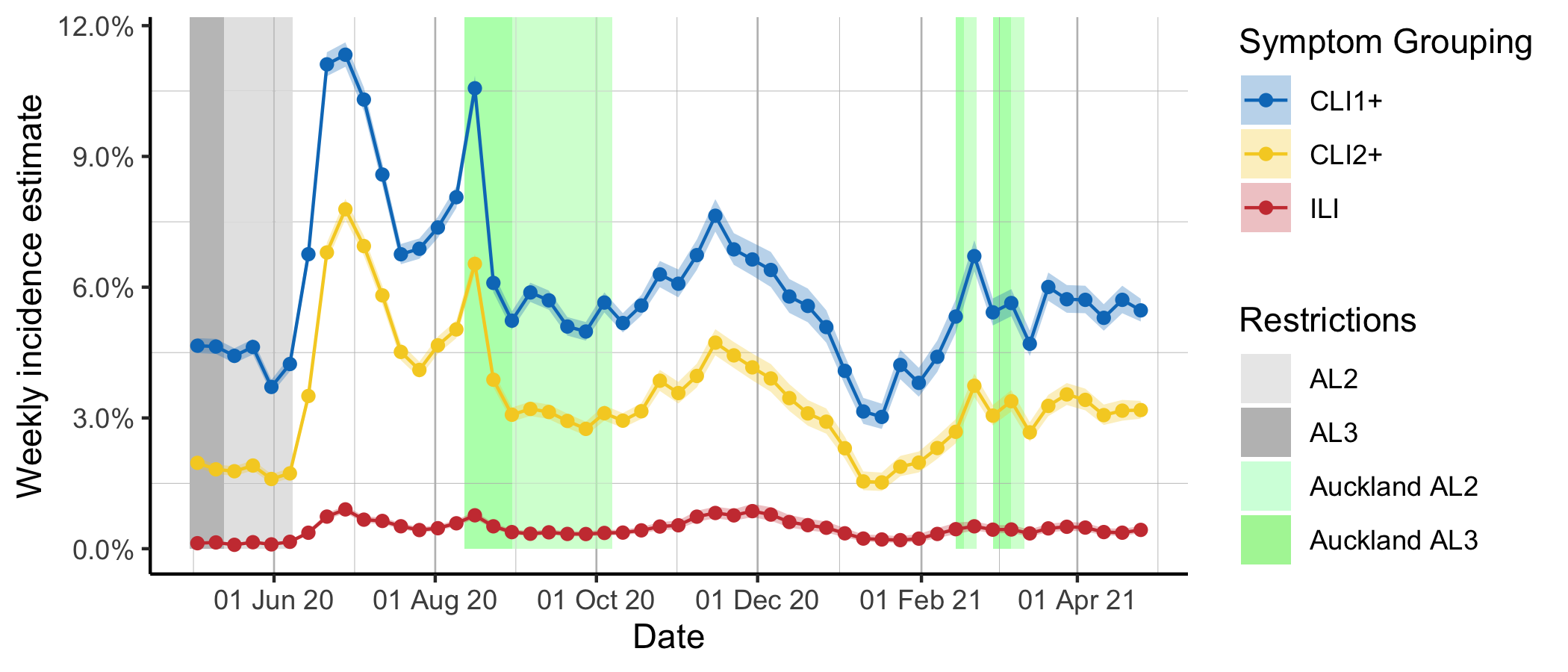}
     \caption{Weekly incidence estimates for the whole of Aotearoa NZ, for three different Symptom Groupings (CLI1+, CLI2+, and ILI) for the year between 27th April 2020 and 25th April 2021 (shaded areas represent 95\% pointwise confidence bands). Alert levels \cite{AlertLevels} (AL2 and AL3 --- AL3 is stronger restrictions than AL2) are shaded. We observe periods of reduced incidence during these elevated Alert Levels periods, as well as some spikes in incidence when the Alert Level is set at 1 (no restrictions), including in July 2020 --- during the winter in Aotearoa New Zealand).
     }
     \label{fig:incidence}
 \end{figure}

Figure \ref{fig:incidence} presents the weekly incidence estimates for the whole of Aotearoa New Zealand and for the CLI1+, CLI2+, and ILI Symptom Groupings after adjusting for consistent responses and age weighting. Consistent responses were weighted by age at a national level, as per Section \ref{sec:age_weighting}. Survey week was the only variable used to explain incidence for each Symptom Grouping. The effect of elevated Alert Level interventions (indicated on the chart as shaded regions) are clearly seen in the reduction of symptom incidence for all Symptom Groupings. That is, the transmission reduction measure put in place to prevent the spread of SARS-CoV-2 also reduced the incidence of other symptoms for respiratory illnesses in general. 

 
\begin{figure}[htb!]
 \centering
  \includegraphics[width=0.95\textwidth]{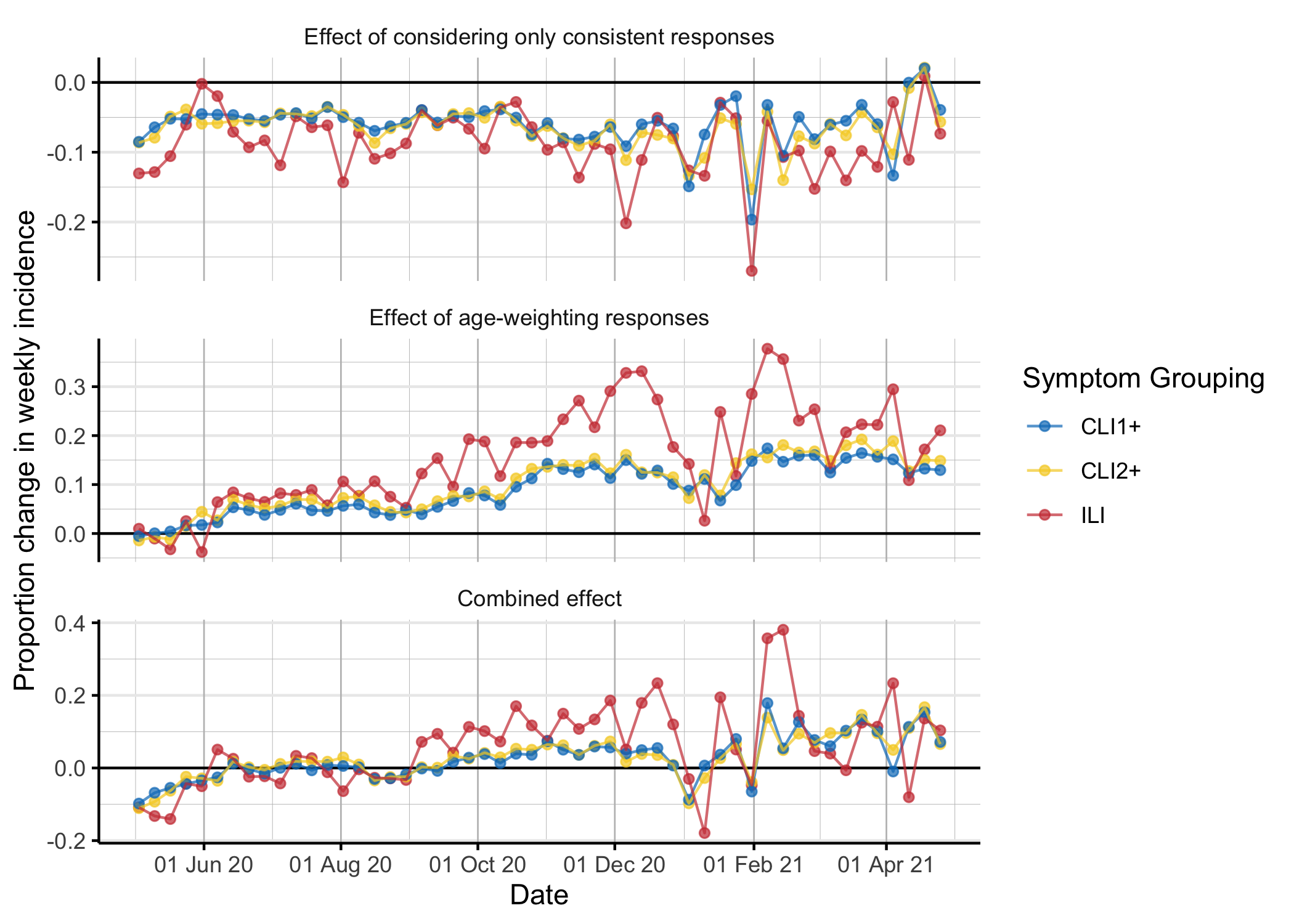}
 \caption{Change in weekly incidence estimates, due to adjusting for symptom reporting bias by considering only `consistent responses' (top), adjusting using age re-weighting (middle), and adjusting for both (bottom), relative to baseline (naive) weekly incidence estimates where all responses are used and there is no age re-weighting.
 }
 \label{fig:effect_of_weighting_and_consistent}
\end{figure}

Figure \ref{fig:effect_of_weighting_and_consistent} shows the effect of adjustments to the incidence estimates in Figure \ref{fig:incidence} for consistent responses and age weighting. Adjusting incidence estimates to include only the population of consistent responses shows an almost universal reduction in incidence rates, by around 5\%. This is due to the removal of respondents who are preferentially (or exclusively) completing the survey when symptomatic, as has been observed in previous analysis \cite{Liu2020}. This effect decreased in February--April of 2021, and coincides with a very stable population of Flutracking survey respondents in this period.


\begin{figure}[htb!]
     \centering
     \includegraphics[width=0.95\textwidth]{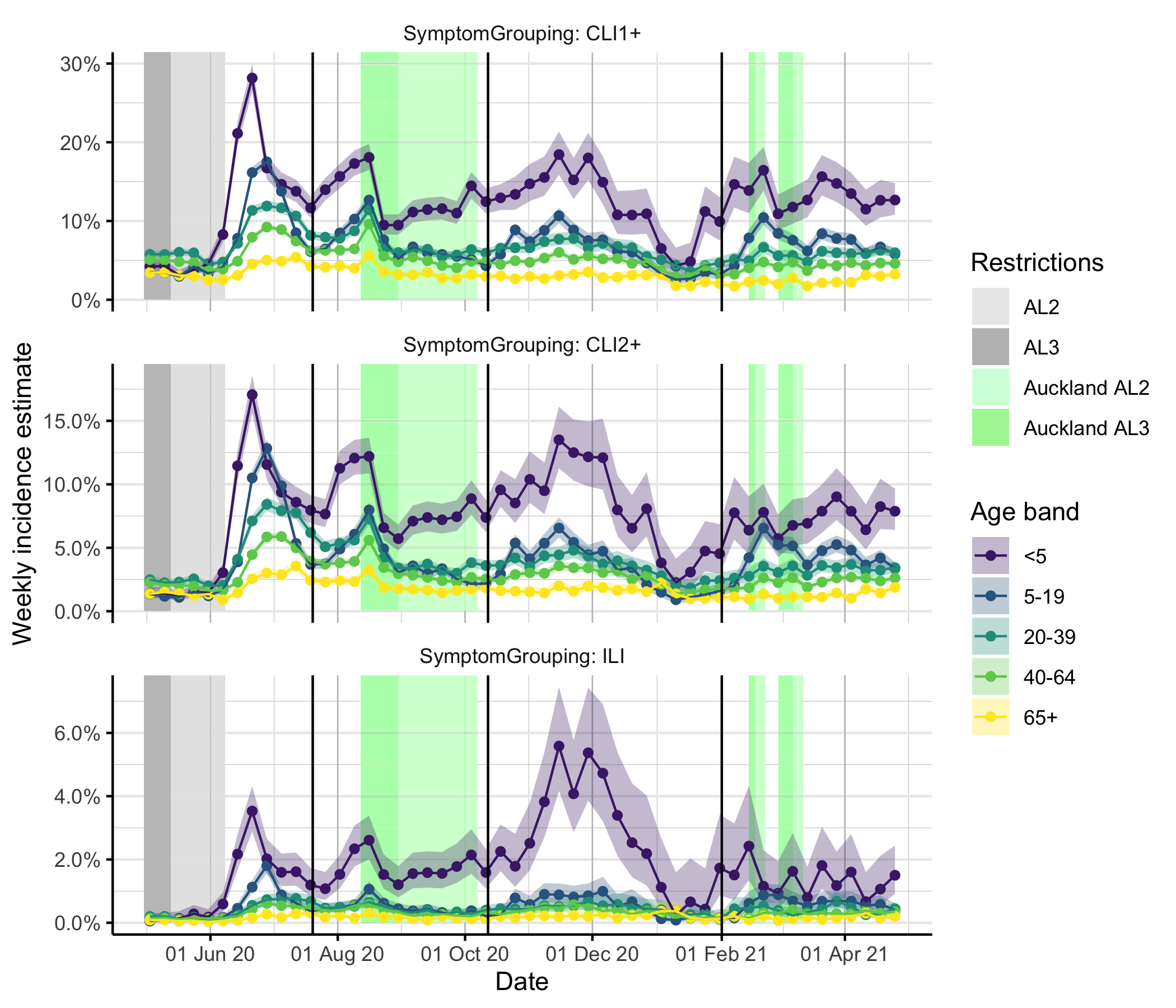}
     \caption{Weekly incidence estimates by selected age bands, for the whole of Aotearoa New Zealand, for the CLI1+, CLI2+, and ILI Symptom Groupings (shaded areas represent 95\% pointwise confidence bands). Alert levels (AL2 and AL3) are indicated by shaded regions. Black lines indicate the beginning of school term periods. 
     }
     \label{fig:incidence_age_annual}
 \end{figure}

When (national level) age-weightings are applied to the incidence estimates, we see an increase in the calculated incidence rate. This is due to younger participants being under-represented in Flutracking responses, relative to the reference population. Younger participants also tend to have higher rates of symptom incidence (Figure~\ref{fig:incidence_age_annual}), hence up-weighting their responses to follow those of the underlying population increases the overall symptom incidence estimates. The size of the effect over the survey period is shown in Figure \ref{fig:effect_of_weighting_and_consistent}.
The size of this effect is not constant over time. The initial half of the survey period saw a higher number of responses from younger participants. When responses from younger participants fell in the latter half of the survey period, the effect due to age-weighted adjustment increased. This is most pronounced in the relative change in incidence rate for ILI symptoms in the age-weighted effect shown in Figure \ref{fig:effect_of_weighting_and_consistent} (middle panel), due to the much higher incidence of ILI symptoms in younger age groups.

Finally, we note that while the relative adjustments for the CLI1+ and CLI2+ incidence rates are mostly smoothly varying over time, the weighting factors applicable to the ILI Symptom Grouping are more volatile due to low total rates of ILI symptoms in Aotearoa New Zealand from early 2020 onward.
 
Figure~\ref{fig:incidence_age_annual} shows the incidence of CLI1+, CLI2+, and ILI symptoms, broken down by age. Responses from consistent responses were weighted by age at a national level, per Section \ref{sec:age_weighting}. Survey week, age band and the interaction effects between the two were used to explain incidence for each Symptom Grouping. The higher rates of incidence for younger age bands is strongest for ILI symptoms. Under-fives have 1.5--2 times the incidence rate of the next highest incidence age band for the CLI2+ Symptom Grouping, and 3--4 times higher for ILI symptoms.


\begin{figure}[htb!]
    \centering
     \includegraphics[width=0.95\textwidth]{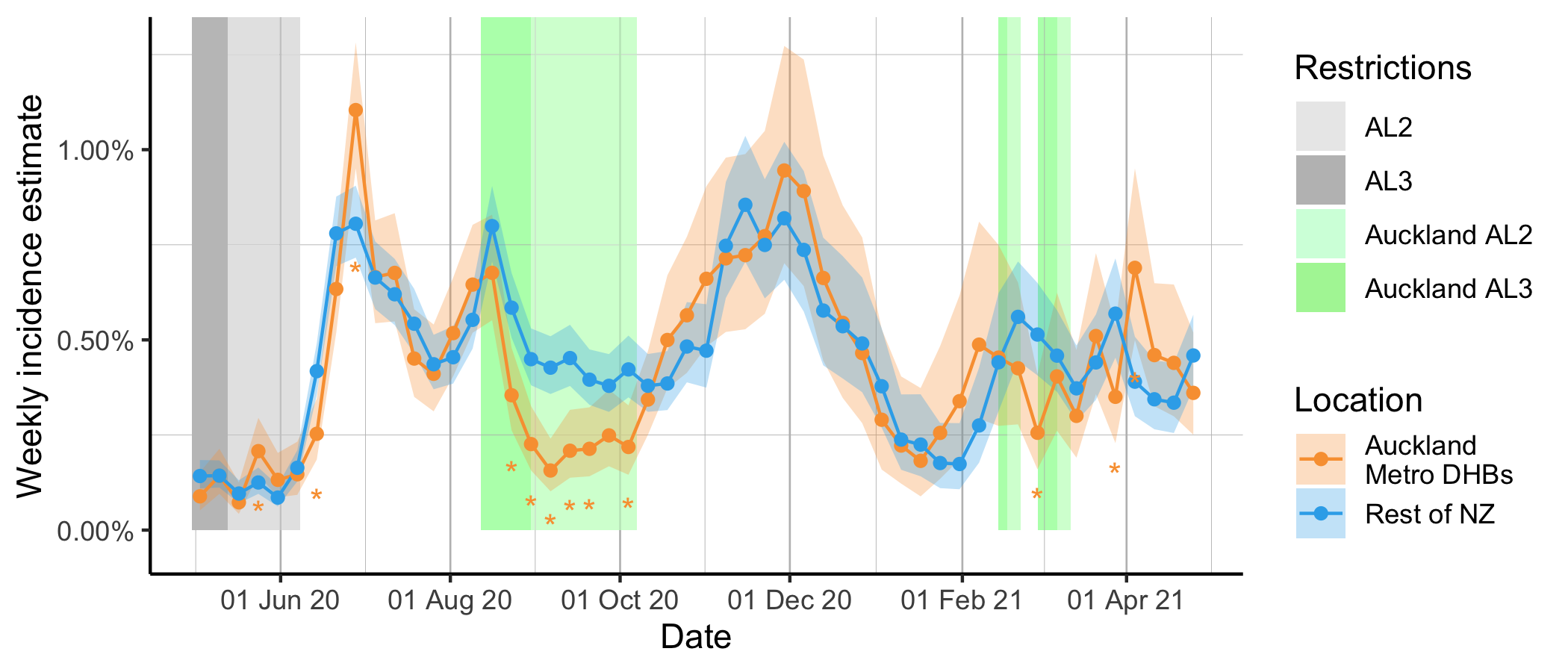}
     \caption{Weekly incidence estimates for the ILI Symptom Grouping split by location (Auckland Metro DHBs and Rest of New Zealand, shaded areas represent 95\% pointwise confidence bands). Stars indicate weeks when the difference in incidence estimate is statistically significant for the Auckland Metro DHBs, relative to the Rest of New Zealand. Alert levels (AL2 and AL3) are indicated by shaded regions.}
     \label{fig:incidence_location}
 \end{figure}

Figure~\ref{fig:incidence_location} shows weekly incidence estimates for ILI split by location for Auckland Metro DHBs and the Rest of New Zealand. Consistent responses were weighted by age using the age distributions in the Auckland Metro DHBs and Rest of New Zealand separately, as per Section \ref{sec:age_weighting}. Survey week, location and the interaction effects between the two were used to explain incidence for ILI. Alert Level interventions designed to prevent the spread of SARS-CoV-2 \cite{AlertLevels}, including those that were principally applied to Auckland alone, are indicated by shaded regions. The effect of these regional interventions is clearly seen in the difference in incidence estimates. The Auckland Metro DHBs have much lower incidence numbers during periods of elevated Alert Levels (stronger restrictions) relative to the rest of the country. 

 
\section{Conclusion}
In this article, we have described an approach for calculating incidence estimates and corresponding confidence intervals for longitudinal participatory health survey data, such as that produced by \emph{Flutracking}. Our method allows for analysis of arbitrary Symptom Groupings --- including two Symptom Groupings developed here for relevance to the COVID-19 response in Aotearoa New Zealand. Our method has several advantages over existing methods for analysis of similar data; namely that it defines a subset of consistent responses for use in calculating incidence estimates; and that it uses re-weighting of responses by participant sub-group in order to account for registration bias. Both of these adjustments increase the rigour and accuracy of the resulting incidence calculations, compared with methods in current use.
We have demonstrated our method and quantified the effect of our methodological improvements by applying it to Flutracking data for Aotearoa New Zealand over a period where symptom incidence was affected by a range of COVID-related behavioural changes and government interventions.



\clearpage
\section*{Acknowledgements}

We wish to thank New Zealand Flutracking, a collaborative initiative between the Ministry of Health (NZ) and Hunter New England Health (NSW, Australia), for the provision of Flutracking data. 
Professor Thomas Lumley provided advice on use of the \texttt{survey} package and the COVID Modelling Aotearoa internal review panel, lead by Dr Matthew Parry, provided valuable feedback on this article.

\section*{Declaration of competing interests}

The authors declare that they have no known competing financial interests or personal relationships that could have appeared to influence the work reported in this paper. 

\section{Supplementary material}


R code and supporting example data that demonstrates the application of this methodology can be found at  \url{https://gitlab.com/cma-public-projects/flutracking-methods-article}\cite{gitlab_repo}. The Rmarkdown document contained within this repository, `\texttt{FluTrackingMethods\_MethodsX.Rmd}', was used to generate the figures found in Section \ref{ssec:method_demo}. This document can be used with a simplified masked dataset to generate the figures in this paper for a 4 month survey period and for responses from Auckland Metro DHBs and Wellington Region DHBs. To obtain the full Aotearoa New Zealand dataset for Flutracking please contact the Ministry of Health (NZ) at \href{mailto:nzmoh_flutracking@health.govt.nz}{\texttt{nzmoh\_flutracking@health.govt.nz}}.

\section{Additional Information}


\subsection{Demographic details of Flutracking participants and responses}
\label{sec:demographic}

Here we present the demographic characteristics of the Flutracking survey population and compare these to the Aotearoa New Zealand population.

Table \ref{tab:demographic_summary} provides a summary of the demographic characteristics of the Flutracking respondents for the period 27th of April, 2020 to 25th of April, 2021. This table lists the average weekly number of responses,  and the number of unique participants, for each demographic group, split by age, ethnicity, gender and region; and for both total participants and for only consistent responses.  

When compared against the demographic characteristics of the Aotearoa New Zealand population (as indicated by the 2018 Census \cite{nz.stat}), we can see that the Flutracking population over-represents individuals who are 40--69 years olds, P\={a}keha/NZ European ethnicity, female, or from Wellington. It particularly under-represents 20--29 year-olds and M\={a}ori and Pacific Peoples. 

When considering only those participants who have had their responses classified as consistent responses at least once, we do not see a substantial change in representation relative to total respondents across any of the individual demographic factors. However, we do find that those groups that are under-represented become even more under-represented when filtering to include only consistent responses.

{\footnotesize
\begin{center}
\begin{longtable}{lp{1.6cm}rrrrr}
\caption{\small Demographic makeup of the Flutracking cohort over a year  (27th April 2020 -- 25th April 2021) as compared to the Aotearoa New Zealand population according to the 2018 Census \cite{nz.stat}. 
Note: the Aotearoa New Zealand level Ethnicity population data from Census 2018 uses the percentage of total responses, whereas the Flutracking data uses Prioritised Ethnicity. The count and percentages of \textit{Responses} are an average of the responses given per group across all survey weeks.  \%$^a$ is the percentage of all responses across the whole year.  $n^b$ is the average weekly number of responses. \textit{All participants} counts the number (and percentage) of unique participants per group in the sample year of data. \textit{Consistent participants} counts the number (and percentage) of unique participants who were considered `consistent' in at least one week of the year.}
\label{tab:demographic_summary} \\

\hline \multicolumn{1}{l}{\textbf{}} & \multicolumn{1}{l}{\textbf{}} & 
\multicolumn{1}{p{2cm}}{\textbf{Average weekly responses \%$^a$ ($n^b$) }} &
\multicolumn{1}{p{2cm}}{\textbf{Average weekly consistent responses \%$^a$  ($n^b$) }} &
\multicolumn{1}{p{2.4cm}}{\textbf{All participants \%  ($n$) }} &
\multicolumn{1}{p{2.2cm}}{\textbf{Consistent participants \%  ($n$) }} &
\multicolumn{1}{p{2cm}}{\textbf{Aotearoa New Zealand Census 2018 \% }} \\ \hline 
\endfirsthead

\multicolumn{7}{c}%
{ \tablename\ \thetable{} -- continued from previous page} \\

\hline \multicolumn{1}{l}{\textbf{}} & \multicolumn{1}{l}{\textbf{}} & 
\multicolumn{1}{p{2cm}}{\textbf{Average weekly responses \%$^a$ ($n^b$) }} &
\multicolumn{1}{p{2cm}}{\textbf{Average weekly consistent responses \%$^a$  ($n^b$) }} &
\multicolumn{1}{p{2.4cm}}{\textbf{All participants \%  ($n$) }} &
\multicolumn{1}{p{2.2cm}}{\textbf{Consistent participants \%  ($n$) }} &
\multicolumn{1}{p{2cm}}{\textbf{Aotearoa New Zealand Census 2018 \% }} \\ \hline 
\endhead

\hline \multicolumn{7}{r}{{Continued on next page}} \\ \hline
\endfoot

\hline \multicolumn{7}{r}{{}}
\endlastfoot
Age (years) & 0-9 & 8.8\% (3924) & 8.6\% (3510) & 10.8\% (9291) & 10.5\% (8203) & 13.1\% \\
& 10-19 & 10.6\% (4736) & 10.5\% (4271) & 11.6\% (9993) & 11.6\% (9036) & 12.9\% \\
& 20-29 & 6.3\% (2801) & 6.1\% (2507) & 8.0\% (6887) & 7.6\% (5933) & 14.1\% \\
& 30-39 & 11.9\% (5294) & 11.7\% (4776) & 13.7\% (11799) & 13.3\% (10390) & 13.0\% \\
& 40-49 & 16.9\% (7540) & 16.8\% (6866) & 17.2\% (14815) & 17.5\% (13611) & 13.0\% \\
& 50-59 & 18.1\% (8080) & 18.3\% (7467) & 16.2\% (13908) & 16.6\% (12946) & 13.0\% \\
& 60-69 & 16.8\% (7472) & 17.1\% (6967) & 13.9\% (11907) & 14.2\% (11051) & 10.4\% \\
& 70-79 & 9.0\% (4026) & 9.2\% (3765) & 7.3\% (6268) & 7.3\% (5718) & 6.7\% \\
& 80+ & 1.5\% (674) & 1.5\% (628) & 1.3\% (1099) & 1.3\% (992) & 3.6\% \\ 
& Total & (44547) & (40756) & (85967) & (77880) & \\ 
\hline
Ethnicity & M\={a}ori & 6.6\% (2931) & 6.5\% (2633) & 7.8\% (6720) & 7.6\% (5933) & 14.7\% \\
& Pacific& 1.5\% (658) & 1.5\% (591) & 1.8\% (1567) & 1.8\% (1393) & 7.2\% \\
& Asian & 3.6\% (1603) & 3.5\% (1434) & 4.3\% (3670) & 4.0\% (3093) & 13.4\% \\
& Other & 4.9\% (2176) & 4.9\% (1980) & 5.2\% (4441) & 5.1\% (3949) & 2.4\% \\
& P\={a}keha/NZ European & 83.5\% (37178) & 83.7\% (34119) & 80.9\% (69569) & 81.6\% (63512) & 62.3\% \\ 
\hline
Gender & Female & 55.7\% (24805)  & 55.7\% (22698)  & 55.5\% (47687) & 55.4\% (43153) & 50.6\%  \\
 & Male & 44.1\% (19665) & 44.1\% (17989)  & 44.3\% (38124)   & 44.4\% (34585)& 49.4\%  \\
 & Other & 0.2\% (77)  & 0.2\% (70) & 0.2\% (156)  & 0.2\% (142) & No data \\ 
 \hline
Region & Northland & 2.8\% (1268) & 2.9\% (1165) & 2.9\% (2466) & 2.9\% (2222) & 3.8\%  \\
 & Auckland & 28.7\% (12762) & 28.5\% (11630) & 29.2\% (25168) &29.0\% (22612)& 33.4\%  \\
 & Waikato & 7.4\% (3312) & 7.4\% (3022) & 7.6\% (6500) &7.6\% (5941) & 9.8\% \\
 & Bay of Plenty & 4.3\% (1921) & 4.3\% (1747) & 4.5\% (3892) &4.4\% (3452)& 6.6\% \\
 & Taranaki & 2.1\% (930) & 2.1\% (843)& 2.1\% (1804)&2.10\% (1615)& 2.5\%\\
 & Gisborne & 0.4\% (176)& 0.4\% (160)& 0.4\% (380) &0.4\% (338)& 1.0\% \\
 & Hawke's Bay & 3.2\% (1433) & 3.2\% (1322) & 3.2\% (2740)&3.20\% (2518)& 3.5\% \\
 & Manawat\={u}-Whanganui & 3.8\% (1692) & 3.8\% (1549)& 3.8\% (3283) &3.9\% (3006)& 5.1\% \\
 & Wellington & 21.9\% (9737) & 22\% (8950)& 20.9\% (17988) &21.0\% (16382)& 10.8\% \\
 & Marlborough & 0.7\% (321) & 0.7\% (293) & 0.7\% (602) &0.7\% (547)& 1.0\% \\
 & Nelson & 1.4\% (614) & 1.4\% (559) & 1.4\% (1172) &1.4\% (1067)& 1.1\% \\
 & Tasman & 1.2\% (529) & 1.2\% (485)& 1.2\% (1047) &1.2\% (950)& 1.1\% \\
 & West Coast & 0.5\% (234) & 0.5\% (215)& 0.6\% (472) &0.5\% (414)& 0.7\% \\
 & Canterbury & 14.2\% (6324) & 14.3\% (5809)& 14.1\% (12098) &14.2\% (11066) & 12.8\% \\
 & Otago & 5.9\% (2626)& 5.9\% (2401) & 5.8\% (4995) &5.8\% (4524)& 4.8\% \\
 & Southland & 1.5\% (660) & 1.5\% (600)& 1.6\% (1347)&1.6\% (1213)& 2.1\%
\end{longtable}
\end{center}
}

\begin{figure}[bht!]
    \centering
    \includegraphics[width=0.75\textwidth]{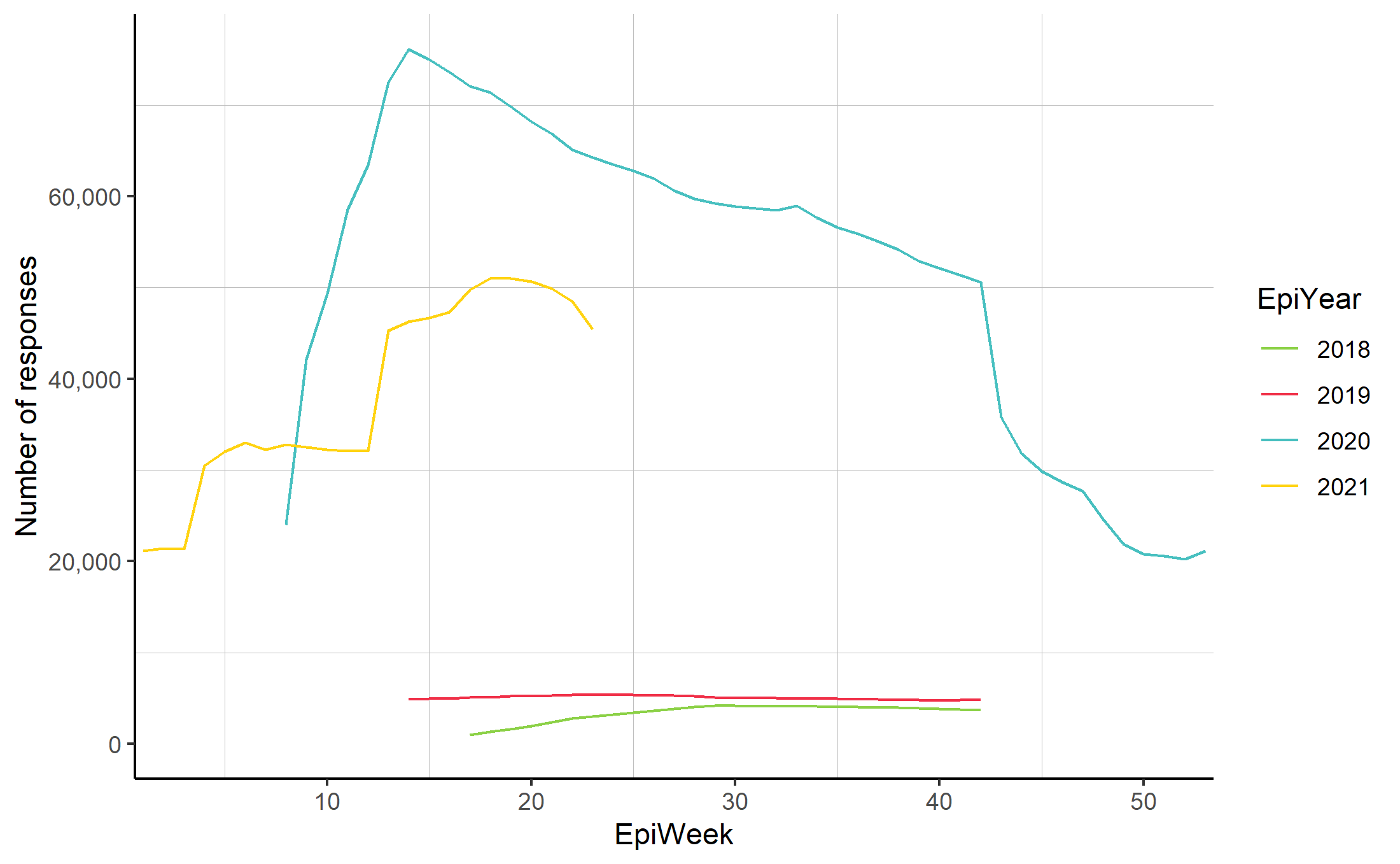}
    \caption{Number of Flutracking responses by epidemiological week (EpiWeek) from 2018 to 2021. We observe that the number of responses is substantially higher for 2020 and 2021 compared to the years prior to the COVID-19 pandemic. Note: in this paper we consider the 12 months from EpiWeek 18 in 2020, to EpiWeek 16 in 2021, inclusive.}
    \label{fig:respondents_summary}
\end{figure}

Figure \ref{fig:respondents_summary} shows the number of responses to Flutracking surveys from 2018 to April 2021. In 2018 and 2019, Flutracking New Zealand had around 4000 -- 5000 participants, with little variation over the May -- October period in which the survey was conducted. In early 2020, the survey was promoted by both Aotearoa New Zealand's Prime Minister and Director General of Health, resulting in a major increase in respondents, peaking at 76,000 \cite{flutracking_press}. This number steadily decreased over the year until October, when respondents were sent an email with the option to pause their responses until the end of summer (February 2021) or the start of the next  `flu' season (April 2021). As a result, responses dropped to just over 20,000, before former respondents were prompted to resume again at several intervals in early 2021. 

The average number of weekly responses over the period 27th April 2020 -- 25th April 2021 was 44,575 responses from a population of 85,967 unique participants. That is, just over half of all registered Flutracking participants respond in a given week; though this number varies over the course of the year, as shown in Figure \ref{fig:respondents_summary}.

\begin{figure}[htb!]
    \centering
    \includegraphics[width=0.75\textwidth]{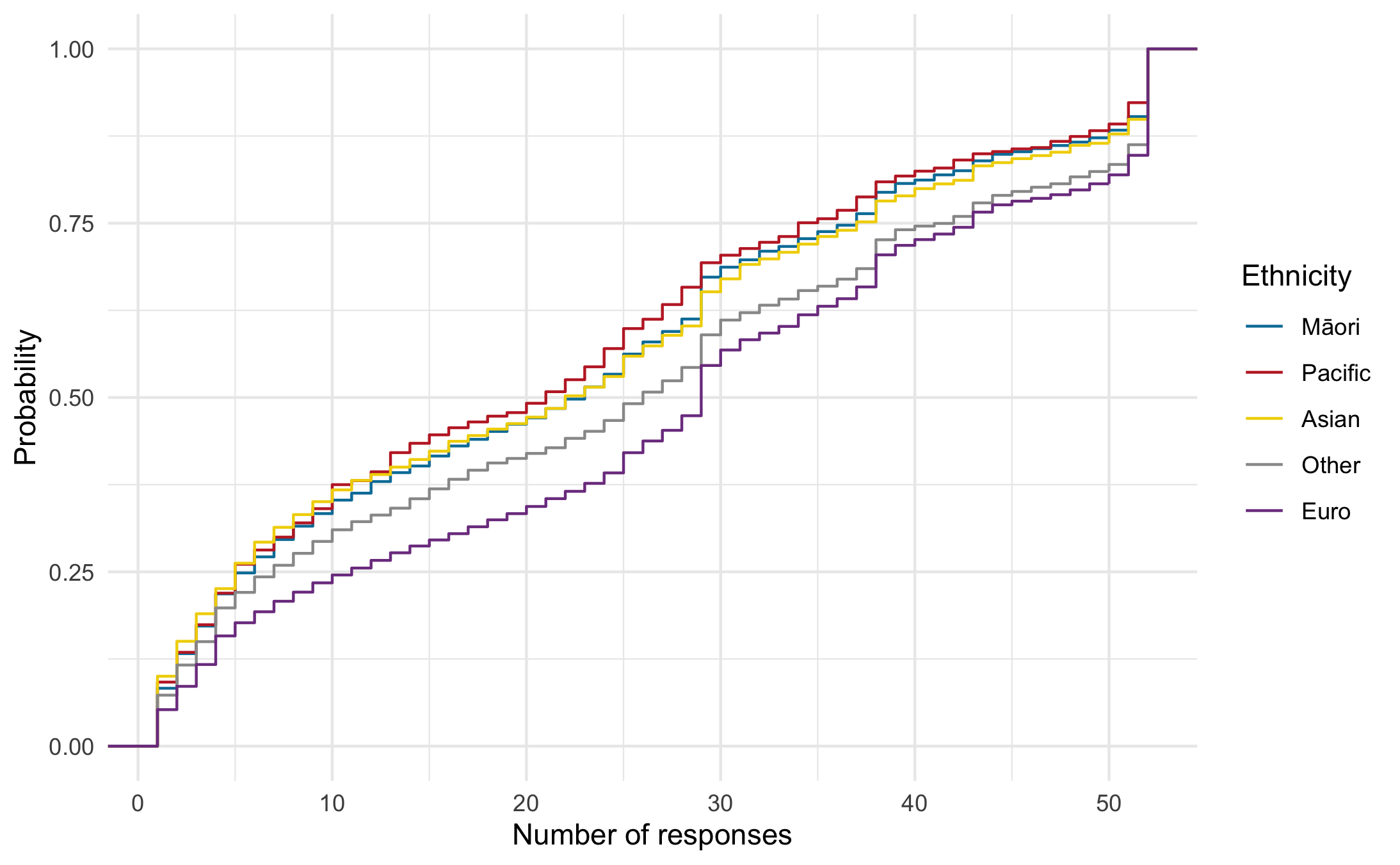}
    \caption{Cumulative probability of a Flutracking respondent having responded to $x$ or fewer surveys over the year between 27th April 2020 and 25th April 2021 by prioritised ethnicity.}
    \label{fig:responses_per_person}
\end{figure}

While similar online health surveillance platforms report that the average participant fills out a low number of surveys over the period of interest \cite{Smolinski2017}, this is not the case for Flutracking. Figure \ref{fig:responses_per_person} shows that the most common number of responses over the course of a year is 52, with around 20\% of respondents never missing a week. This may be aided by respondents' ability to submit surveys up to 4 weeks in arrears. The spikes around 28 and 38 weeks worth of responses are likely to correspond to the emails encouraging people to opt in or out for the `flu' season, as mentioned above.

\subsection{Selecting the criteria for consistent responses}

\begin{figure}[htb!]
    \centering
    \includegraphics[width=0.95\textwidth]{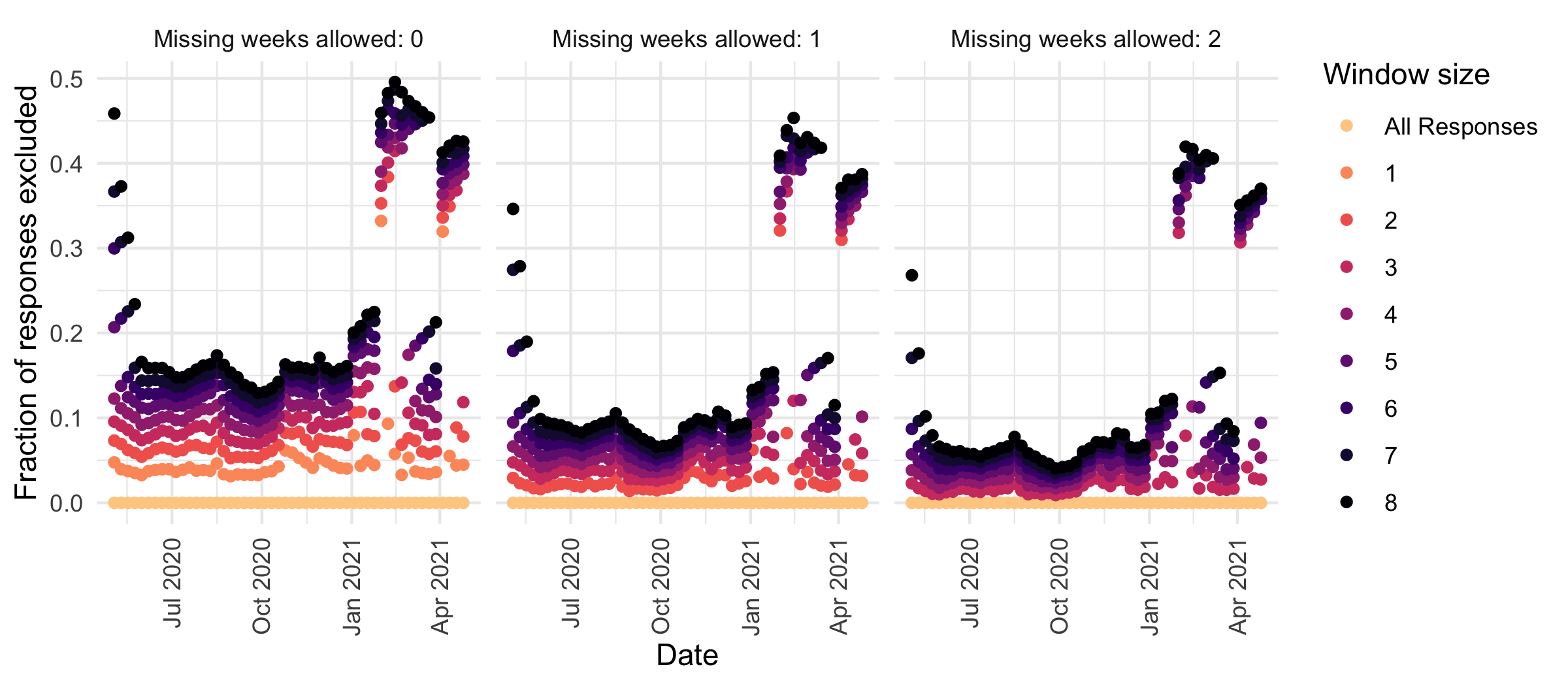}
    \includegraphics[width=0.95\textwidth]{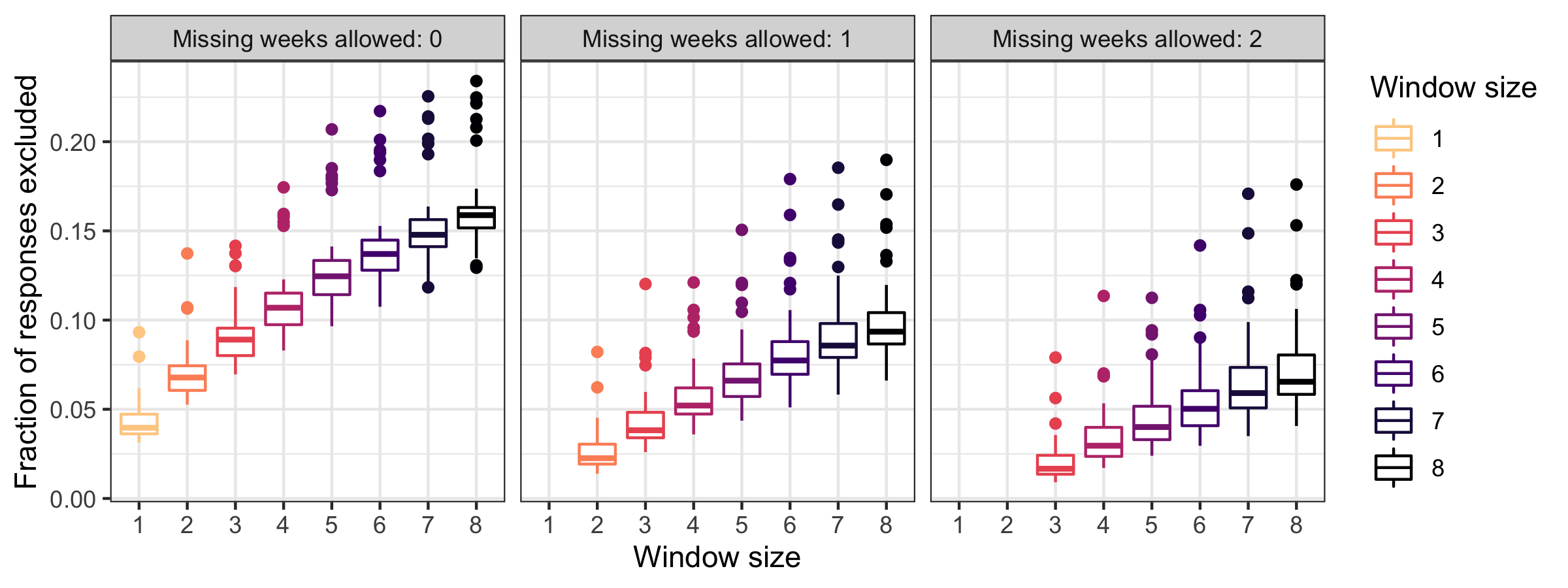}
    \caption{The upper plot shows the impact of `Window size' and `Missing weeks allowed' on the fraction of responses excluded in each week of the year. Here we see the large spikes around February and April, due to the survey emails being paused over the summer break until these times for a large proportion of participants. The lower plot filters out any weeks with over 25\% of responses excluded, and produces a box plot to summarise the impact for each `Window size' and `Missing weeks allowed'.}
    \label{fig:responsesChange}
 \end{figure}

\begin{figure}[htb!]
    \centering
    \includegraphics[width=0.9\textwidth]{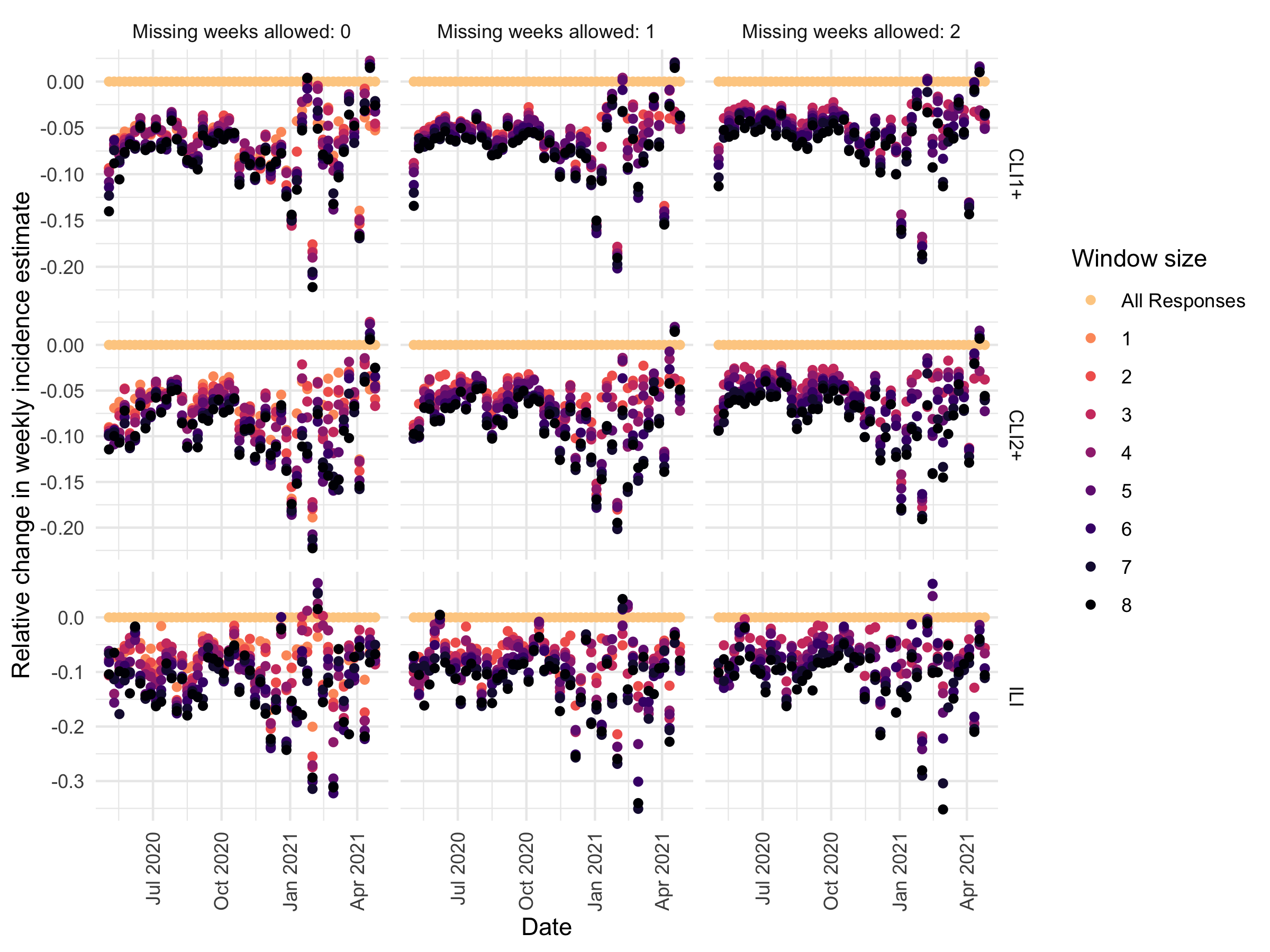}
    \includegraphics[width=0.9\textwidth]{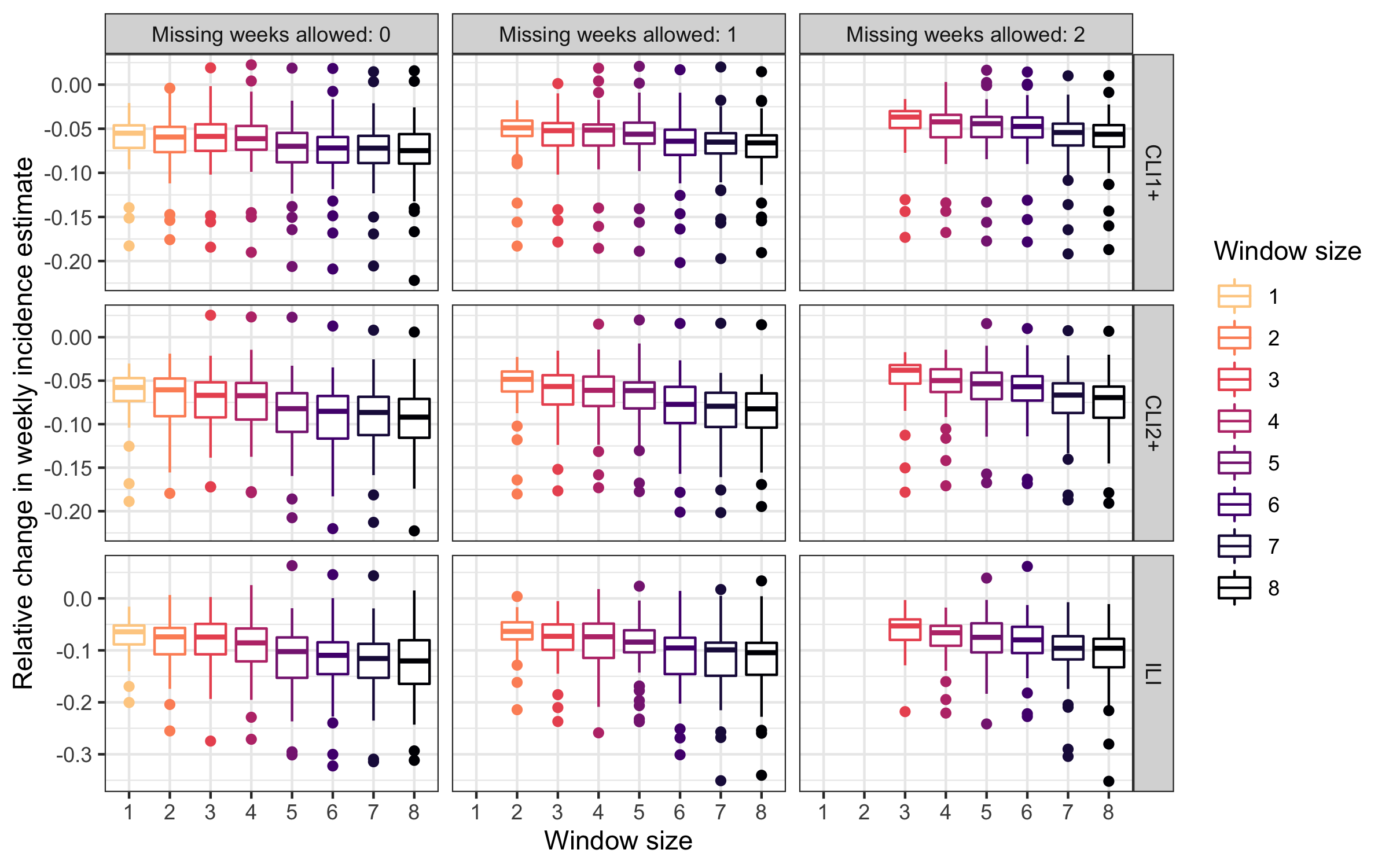}
    \caption{The upper plot shows the relative change in weekly incidence estimate for a range of `Window sizes' and `Missing weeks allowed' for the three Symptom GroupingsL CLI1+, CLI2+, and ILI, for each week of the year. The lower plot presents a box plot of the above results to summarise the impact for each `Window size' and `Missing weeks allowed' over the whole year.}
    \label{fig:incidenceChange}
 \end{figure}

In Figure~\ref{fig:Consistent_tradeoffs} we used medians to summarise the impact of the choice of `Window size' and `Missing weeks allowed' on the  fraction of responses excluded each week and the relative change in weekly incidence estimates. The magnitude of these effects varies throughout the year. Here we show the impact of window size and missing weeks allowed on the number of consistent responses (Figure~\ref{fig:responsesChange}) and the estimated incidence rate (Figure~\ref{fig:incidenceChange}) throughout the year.

\clearpage
 %

\begin{thebibliography}{10}
\expandafter\ifx\csname url\endcsname\relax
  \def\url#1{\texttt{#1}}\fi
\expandafter\ifx\csname urlprefix\endcsname\relax\def\urlprefix{URL }\fi
\expandafter\ifx\csname href\endcsname\relax
  \def\href#1#2{#2} \def\path#1{#1}\fi

\bibitem{survey_package_r}
T.~Lumley, survey: analysis of complex survey samples, r package version 4.0
  (2020).

\bibitem{r_core_team}
{R Core Team}, \href{https://www.R-project.org/}{R: A Language and Environment
  for Statistical Computing}, R Foundation for Statistical Computing, Vienna,
  Austria (2021).
\newline\urlprefix\url{https://www.R-project.org/}

\bibitem{flutracking_info}
{Flutracking}.net, \url{https://info.flutracking.net/about/}.

\bibitem{dalton2009flutracking}
C.~Dalton, D.~Durrheim, J.~Fejsa, L.~Francis, S.~Carlson, E.~T. d'Espaignet,
  F.~Tuyl, Flutracking: a weekly {A}ustralian community online survey of
  influenza-like illness in 2006, 2007 and 2008, Communicable diseases
  intelligence quarterly report 33~(3) (2009) 316--322.

\bibitem{Baltrusaitis2018}
K.~Baltrusaitis, K.~Noddin, C.~Nguyen, A.~Crawley, J.~Brownstein, L.~White,
  Evaluation of approaches that adjust for biases in participatory surveillance
  systems, Online Journal of Public Health Informatics 10 (05 2018).
\newblock \href {https://doi.org/10.5210/ojphi.v10i1.8908}
  {\path{doi:10.5210/ojphi.v10i1.8908}}.

\bibitem{Liu2020}
D.~Liu, L.~Mitchell, R.~C. Cope, S.~J. Carlson, J.~V. Ross,
  \href{https://www.sciencedirect.com/science/article/pii/S175543652030030X}{Elucidating
  user behaviours in a digital health surveillance system to correct prevalence
  estimates}, Epidemics 33 (2020) 100404.
\newblock \href {https://doi.org/10.1016/j.epidem.2020.100404}
  {\path{doi:10.1016/j.epidem.2020.100404}}.
\newline\urlprefix\url{https://www.sciencedirect.com/science/article/pii/S175543652030030X}

\bibitem{Friesema2009_Internet_based_monitoring}
I.~Friesema, C.~Koppeschaar, G.~Donker, F.~Dijkstra, S.~van Noort,
  R.~Smallenburg, W.~van~der Hoek, M.~van~der Sande, Internet-based monitoring
  of influenza-like illness in the general population: Experience of five
  influenza seasons in the netherlands, Vaccine 27~(45) (2009) 6353--6357.
\newblock \href {https://doi.org/10.1016/j.vaccine.2009.05.042}
  {\path{doi:10.1016/j.vaccine.2009.05.042}}.

\bibitem{Debin2013}
M.~Debin, C.~Turbelin, T.~Blanchon, I.~Bonmarin, A.~Falchi, T.~Hanslik,
  D.~Levy-Bruhl, C.~Poletto, V.~Colizza,
  \href{https://pubmed.ncbi.nlm.nih.gov/24040020}{Evaluating the feasibility
  and participants' representativeness of an online nationwide surveillance
  system for influenza in {F}rance}, {PLoS} One 8~(9) (2013) e73675--e73675.
\newblock \href {https://doi.org/10.1371/journal.pone.0073675}
  {\path{doi:10.1371/journal.pone.0073675}}.
\newline\urlprefix\url{https://pubmed.ncbi.nlm.nih.gov/24040020}

\bibitem{Scarpino2020_Socioeconomic_bias_influenza}
S.~V. Scarpino, J.~G. Scott, R.~M. Eggo, B.~Clements, N.~B. Dimitrov, L.~A.
  Meyers, Socioeconomic bias in influenza surveillance, {PLOS} Computational
  Biology 16~(7) (2020) e1007941.
\newblock \href {https://doi.org/10.1371/journal.pcbi.1007941}
  {\path{doi:10.1371/journal.pcbi.1007941}}.

\bibitem{Cantarelli2014}
P.~Cantarelli, M.~Debin, C.~Turbelin, C.~Poletto, T.~Blanchon, A.~Falchi,
  T.~Hanslik, I.~Bonmarin, D.~Levy-Bruhl, A.~Micheletti, D.~Paolotti,
  A.~Vespignani, J.~Edmunds, K.~Eames, R.~Smallenburg, C.~Koppeschaar, A.~O.
  Franco, V.~Faustino, A.~Carnahan, M.~Rehn, V.~Colizza,
  \href{https://pubmed.ncbi.nlm.nih.gov/25240865}{The representativeness of a
  european multi-center network for influenza-like-illness participatory
  surveillance}, BMC public health 14 (2014) 984--984.
\newblock \href {https://doi.org/10.1186/1471-2458-14-984}
  {\path{doi:10.1186/1471-2458-14-984}}.
\newline\urlprefix\url{https://pubmed.ncbi.nlm.nih.gov/25240865}

\bibitem{Brownstein2017_Combining_Participatory_Influenza}
J.~S. Brownstein, S.~Chu, A.~Marathe, M.~V. Marathe, A.~T.~N. BS, D.~Paolotti,
  N.~Perra, D.~P. MS, M.~Santillana, S.~Swarup, M.~Tizzoni, A.~Vespignani,
  A.~K.~S. Vullikanti, M.~L.~W. MA, Q.~Zhang, Combining participatory influenza
  surveillance with modeling and forecasting: Three alternative approaches,
  JMIR Public Health And Surveillance 3~(4) (2017).
\newblock \href {https://doi.org/10.2196/publichealth.7344}
  {\path{doi:10.2196/publichealth.7344}}.

\bibitem{flutracking_june2022}
Flutracking, Weekly interim report {N}ew {Z}ealand - week ending 29 {M}ay 2022,
  https://www.flutracking.net/Info/Report/202222/NZ, accessed: 06-06-2022
  (2022).

\bibitem{Chunara2015}
R.~Chunara, E.~Goldstein, O.~Patterson-Lomba, J.~S. Brownstein,
  \href{https://doi.org/10.1038/srep09540}{Estimating influenza attack rates in
  the {U}nited {S}tates using a participatory cohort}, Scientific Reports 5~(1)
  (2015) 9540.
\newblock \href {https://doi.org/10.1038/srep09540}
  {\path{doi:10.1038/srep09540}}.
\newline\urlprefix\url{https://doi.org/10.1038/srep09540}

\bibitem{Lavezzo2020Vo}
E.~Lavezzo, E.~Franchin, C.~Ciavarella, G.~Cuomo-Dannenburg, L.~Barzon,
  C.~Del~Vecchio, L.~Rossi, R.~Manganelli, A.~Loregian, N.~Navarin, D.~Abate,
  M.~Sciro, S.~Merigliano, E.~De~Canale, M.~C. Vanuzzo, V.~Besutti, F.~Saluzzo,
  F.~Onelia, M.~Pacenti, S.~G. Parisi, G.~Carretta, D.~Donato, L.~Flor,
  S.~Cocchio, G.~Masi, A.~Sperduti, L.~Cattarino, R.~Salvador, M.~Nicoletti,
  F.~Caldart, G.~Castelli, E.~Nieddu, B.~Labella, L.~Fava, M.~Drigo, K.~A.~M.
  Gaythorpe, K.~E.~C. Ainslie, M.~Baguelin, S.~Bhatt, A.~Boonyasiri, O.~Boyd,
  L.~Cattarino, C.~Ciavarella, H.~L. Coupland, Z.~Cucunubá,
  G.~Cuomo-Dannenburg, B.~A. Djafaara, C.~A. Donnelly, I.~Dorigatti, S.~L. van
  Elsland, R.~FitzJohn, S.~Flaxman, K.~A.~M. Gaythorpe, W.~D. Green,
  T.~Hallett, A.~Hamlet, D.~Haw, N.~Imai, B.~Jeffrey, E.~Knock, D.~J. Laydon,
  T.~Mellan, S.~Mishra, G.~Nedjati-Gilani, P.~Nouvellet, L.~C. Okell, K.~V.
  Parag, S.~Riley, H.~A. Thompson, H.~J.~T. Unwin, R.~Verity, M.~A.~C. Vollmer,
  P.~G.~T. Walker, C.~E. Walters, H.~Wang, Y.~Wang, O.~J. Watson, C.~Whittaker,
  L.~K. Whittles, X.~Xi, N.~M. Ferguson, A.~R. Brazzale, S.~Toppo, M.~Trevisan,
  V.~Baldo, C.~A. Donnelly, N.~M. Ferguson, I.~Dorigatti, A.~Crisanti,
  {Imperial College COVID-19 Response Team},
  \href{https://doi.org/10.1038/s41586-020-2488-1}{Suppression of a
  {SARS}-{CoV}-2 outbreak in the {Italian} municipality of {Vo}’}, Nature
  584~(7821) (2020) 425--429.
\newblock \href {https://doi.org/10.1038/s41586-020-2488-1}
  {\path{doi:10.1038/s41586-020-2488-1}}.
\newline\urlprefix\url{https://doi.org/10.1038/s41586-020-2488-1}

\bibitem{Wells2020symptoms}
P.~M. Wells, K.~M. Doores, S.~Couvreur, R.~Martin~Martinez, J.~Seow, C.~Graham,
  S.~Acors, N.~Kouphou, S.~Neil, R.~Tedder, P.~Matos, K.~Poulton,
  M.~Jose~Lista, R.~Dickenson, H.~Sertkaya, T.~Maguire, E.~Scourfield,
  R.~Bowyer, D.~Hart, A.~O{\textquoteright}Byrne, K.~Steele, O.~Hemmings,
  C.~Rosadas, M.~McClure, J.~Capedevila-Pujol, J.~wolf, S.~Ourseilin, M.~Brown,
  M.~Malim, T.~Spector, C.~Steves,
  \href{https://www.medrxiv.org/content/early/2020/07/30/2020.07.29.20162701}{Estimates
  of the rate of infection and asymptomatic {COVID-19} disease in a population
  sample from {SE England}}, medRxiv (2020).
\newblock \href
  {http://arxiv.org/abs/https://www.medrxiv.org/content/early/2020/07/30/2020.07.29.20162701.full.pdf}
  {\path{arXiv:https://www.medrxiv.org/content/early/2020/07/30/2020.07.29.20162701.full.pdf}},
  \href {https://doi.org/10.1101/2020.07.29.20162701}
  {\path{doi:10.1101/2020.07.29.20162701}}.
\newline\urlprefix\url{https://www.medrxiv.org/content/early/2020/07/30/2020.07.29.20162701}

\bibitem{SymptomClusters}
C.~H. Sudre, K.~Lee, M.~Ni~Lochlainn, T.~Varsavsky, B.~Murray, M.~S. Graham,
  C.~Menni, M.~Modat, R.~C. Bowyer, L.~H. Nguyen, D.~A. Drew, A.~D. Joshi,
  W.~Ma, C.~G. Guo, C.~H. Lo, S.~Ganesh, A.~Buwe, J.~Capdevila~Pujol,
  J.~Lavigne~du Cadet, A.~Visconti, M.~Freydin, J.~S. El~Sayed~Moustafa,
  M.~Falchi, R.~Davies, M.~F. Gomez, T.~Fall, M.~J. Cardoso, J.~Wolf, P.~W.
  Franks, A.~T. Chan, T.~D. Spector, C.~J. Steves, S.~Ourselin,
  \href{https://www.medrxiv.org/content/early/2020/06/16/2020.06.12.20129056}{Symptom
  clusters in {Covid19}: A potential clinical prediction tool from the {COVID}
  symptom study app}, medRxiv (2020).
\newblock \href
  {http://arxiv.org/abs/https://www.medrxiv.org/content/early/2020/06/16/2020.06.12.20129056.full.pdf}
  {\path{arXiv:https://www.medrxiv.org/content/early/2020/06/16/2020.06.12.20129056.full.pdf}},
  \href {https://doi.org/10.1101/2020.06.12.20129056}
  {\path{doi:10.1101/2020.06.12.20129056}}.
\newline\urlprefix\url{https://www.medrxiv.org/content/early/2020/06/16/2020.06.12.20129056}

\bibitem{smolinski2015flu}
M.~S. Smolinski, A.~W. Crawley, K.~Baltrusaitis, R.~Chunara, J.~M. Olsen,
  O.~W{\'o}jcik, M.~Santillana, A.~Nguyen, J.~S. Brownstein, Flu near you:
  crowdsourced symptom reporting spanning 2 influenza seasons, American journal
  of public health 105~(10) (2015) 2124--2130.

\bibitem{Paolotti2014}
D.~Paolotti, A.~Carnahan, V.~Colizza, K.~Eames, J.~Edmunds, G.~Gomes,
  C.~Koppeschaar, M.~Rehn, R.~Smallenburg, C.~Turbelin, S.~{Van Noort},
  A.~Vespignani,
  \href{https://www.sciencedirect.com/science/article/pii/S1198743X14601889}{Web-based
  participatory surveillance of infectious diseases: the {Influenzanet}
  participatory surveillance experience}, Clinical Microbiology and Infection
  20~(1) (2014) 17--21.
\newblock \href {https://doi.org/10.1111/1469-0691.12477}
  {\path{doi:10.1111/1469-0691.12477}}.
\newline\urlprefix\url{https://www.sciencedirect.com/science/article/pii/S1198743X14601889}

\bibitem{MoHCovidSymptoms}
{Ministry of Health}, {About COVID-19},
  \url{https://www.health.govt.nz/covid-19-novel-coronavirus/covid-19-health-advice-public/about-covid-19},
  accessed 2021-12-01 (2021).
  
 \bibitem{Kalimeri2019}
K.~Kalimeri, M.~Delfino, C.~ Cattuto, D.~Perrotta, V.~Colizza, C.~Guerrisi, C.~Turbelin, J.~Duggan, J.~Edmunds, C.~Obi, R.~Pebody. Unsupervised extraction of epidemic syndromes from participatory influenza surveillance self-reported symptoms. PLoS computational biology, 15(4) (2019) p.e1006173.

\bibitem{BAKER2012}
M.~G. Baker, L.~T. Barnard, A.~Kvalsvig, A.~Verrall, J.~Zhang, M.~Keall,
  N.~Wilson, T.~Wall, P.~Howden-Chapman,
  \href{https://www.sciencedirect.com/science/article/pii/S0140673611617807}{Increasing
  incidence of serious infectious diseases and inequalities in {New Zealand}: a
  national epidemiological study}, The Lancet 379~(9821) (2012) 1112--1119.
\newblock \href {https://doi.org/https://doi.org/10.1016/S0140-6736(11)61780-7}
  {\path{doi:https://doi.org/10.1016/S0140-6736(11)61780-7}}.
\newline\urlprefix\url{https://www.sciencedirect.com/science/article/pii/S0140673611617807}

\bibitem{HealthyHousing}
M.~G. Baker, A.~McDonald, J.~Zhang, P.~Howden-Chapman,
  \href{http://www.healthyhousing.org.nz/wp-content/uploads/2010/01/HH-Crowding-ID-Burden-25-May-2013.pdf}{Infectious
  diseases attributable to household crowding in {New Zealand}: A systematic
  review and burden of disease estimate} (May 2013).
\newline\urlprefix\url{http://www.healthyhousing.org.nz/wp-content/uploads/2010/01/HH-Crowding-ID-Burden-25-May-2013.pdf}

\bibitem{Ingham2019}
T.~Ingham, M.~Keall, B.~Jones, D.~R.~T. Aldridge, A.~C. Dowell, C.~Davies,
  J.~Crane, J.~B. Draper, L.~O. Bailey, H.~Viggers, T.~V. Stanley,
  P.~Leadbitter, M.~Latimer, P.~Howden-Chapman,
  \href{https://doi.org/10.1136/thoraxjnl-2018-212979}{Damp mouldy housing and
  early childhood hospital admissions for acute respiratory infection: a case
  control study}, Thorax 74~(9) (2019) 849--857.
\newblock \href {https://doi.org/10.1136/thoraxjnl-2018-212979}
  {\path{doi:10.1136/thoraxjnl-2018-212979}}.
\newline\urlprefix\url{https://doi.org/10.1136/thoraxjnl-2018-212979}

\bibitem{AsthmaFoundation}
J.~Zhang, L.~Telfar~Barnard,
  \href{https://www.asthmafoundation.org.nz/assets/documents/Respiratory-Impact-report-final-2021Aug11.pdf}{The
  impact of respiratory disease in {N}ew {Z}ealand: 2020 update} (August 2021).
\newline\urlprefix\url{https://www.asthmafoundation.org.nz/assets/documents/Respiratory-Impact-report-final-2021Aug11.pdf}

\bibitem{Dalton2017_Insights_Flutracking_Thirteen}
Dalton, C., Carlson, S., Butler, M., Cassano, D., Clarke, S., Fejsa, J., Durrheim, D., Insights from flutracking: thirteen tips to growing a web-based participatory surveillance system. JMIR Public Health and Surveillance, 3(3) (2017), p.e7333.

\bibitem{Smolinski2017}
M.~S. Smolinski, A.~W. Crawley, J.~M. Olsen, T.~Jayaraman, M.~Libel,
  \href{http://www.ncbi.nlm.nih.gov/pubmed/29021131}{Participatory disease
  surveillance: Engaging communities directly in reporting, monitoring, and
  responding to health threats}, JMIR Public Health Surveillance 3~(4) (2017)
  e62.
\newblock \href {https://doi.org/10.2196/publichealth.7540}
  {\path{doi:10.2196/publichealth.7540}}.
\newline\urlprefix\url{http://www.ncbi.nlm.nih.gov/pubmed/29021131}

\bibitem{StatisticsNZ2021}
{Statistics NZ}, Technical notes {DHB} ethnic projections (2021 update) (2021).

\bibitem{efron_better_1987}
B.~Efron, \href{http://www.jstor.org/stable/2289153}{Better {Bootstrap}
  {Confidence} {Intervals}: {Rejoinder}}, Journal of the American Statistical
  Association 82~(397) (1987) 198--200, publisher: [American Statistical
  Association, Taylor \& Francis, Ltd.].
\newblock \href {https://doi.org/10.2307/2289153} {\path{doi:10.2307/2289153}}.
\newline\urlprefix\url{http://www.jstor.org/stable/2289153}

\bibitem{AlertLevels}
N.~Z. Government, {COVID-19} alert system in {N}ew {Z}ealand,
  \url{https://covid19.govt.nz/alert-system/alert-system-overview/}, accessed 2020-08-29 (2020).

\bibitem{ggplot2}
Wickham, H., 2011. ggplot2. Wiley {I}nterdisciplinary {R}eviews: computational statistics, 3(2), pp.180-185.


\bibitem{gitlab_repo}
FluTracking Data Analysis,
COVID-19 Modelling Aotearoa Public Projects,
\url{https://gitlab.com/cma-public-projects/fluTracking-data-analysis}

\bibitem{nz.stat}
National population projections, by age and sex, 2020(base)-2073 information on
  table., \url{http://nzdotstat.stats.govt.nz/wbos/Index.aspx\#} (2021).

\bibitem{flutracking_press}
A.~Mau, {Coronavirus: Dr Ashley Bloomfield loves it, but what is the flu
  tracker?},
  Stuff.co.nz\url{https://www.stuff.co.nz/national/health/coronavirus/120746580/coronavirus-dr-ashley-bloomfield-loves-it-but-what-is-the-flu-tracker},
  2020-04-02.

\end{thebibliography}





\bibliographystyle{elsarticle-num}

%
%



\end{document}